\RequirePackage{ifpdf}
\ifpdf
    \documentclass[pdftex,envcountsect,envcountsame,orivec]{llncs}
\else
    \documentclass[dvips,envcountsect,envcountsame,orivec]{llncs}
\fi

\usepackage{xspace}
\usepackage{textgreek}
\usepackage{latexsym,amssymb,amsmath,amsfonts}
\usepackage{tabularx}
\usepackage{paralist}
\usepackage{todonotes}
\usepackage{fancyvrb}
\usepackage{listings}
\usepackage{booktabs}

\usepackage{enumerate}
\usepackage[latin1]{inputenc}
\usepackage[pdftitle   = {{BPM}},
            pdfauthor  = {{Calvanese -Ghilardi - Gianola - Montali - Rivkin}},
            pdfsubject = {{}}]{hyperref}
\ifpdf \usepackage{url} \else \usepackage{breakurl} \fi
\usepackage{comment}
\usepackage{graphicx}
\usepackage{tikz}
\usetikzlibrary{matrix,arrows, calc,fit,shapes, trees, positioning, decorations.markings, patterns,calc} 
\usepackage{bpmn-events}
\usepackage{bpmn-gateways}
\usepackage{wrapfig}

\usepackage{amsmath,amssymb}

\usepackage[linesnumberedhidden,ruled,vlined]{algorithm2e}
\usepackage{graphicx}

\title{Formal Modeling and SMT-Based Parameterized Verification of Multi-Case Data-Aware BPMN}

\author{%
Diego Calvanese$^1$ , Silvio Ghilardi$^2$,  Alessandro Gianola$^1$ , \\ Marco Montali$^1$, Andrey Rivkin$^1$ 
}

\institute{%
$^1$Faculty of Computer Science, Free University of Bozen-Bolzano (Italy)\\
$^2$Dipartimento di Matematica, Universit\`a degli Studi di Milano (Italy) 
}

\newcommand{\D}{\ensuremath{\mathcal{D}}}

\newcommand{\I}{\ensuremath{\mathcal{I}}}

\newcommand{\M}{\ensuremath{\mathcal{M}}}

\renewcommand{\P}{\ensuremath{\mathcal{P}}}

\newcommand{\R}{\ensuremath{\mathcal{R}}}
\renewcommand{\S}{\ensuremath{\mathcal{S}}}

\newcommand{\V}{\ensuremath{\mathcal{V}}}

\newcommand{\set}[1]{\{#1\}}                      

\newcommand{\tup}[1]{\langle #1\rangle}            

\newcommand{\inlinetitle}[1]{\smallskip\noindent\textbf{#1.}\xspace}

\newcolumntype{C}{>{\centering\arraybackslash}X}

\makeatletter
\g@addto@macro\normalsize{%
\setlength{\abovecaptionskip}{-2pt}
\setlength{\belowcaptionskip}{-10pt}
\setlength\abovedisplayskip{3pt}
\setlength\belowdisplayskip{3pt}
\setlength\abovedisplayshortskip{3pt}
\setlength\belowdisplayshortskip{3pt}
}
\makeatother

\newcounter{dummy} 
\newcounter{dummy1} 
\newcounter{dummy2}
\newcounter{dummy3} 
\newcounter{dummy4}
\newcounter{dummy5} 
\newcounter{dummy6}

\usepackage[thmmarks,amsmath]{ntheorem}
\theorempreskip{1pt}
\theorempostskip{1pt}

\theoremseparator{.}
\theorembodyfont{\itshape}
\theoremsymbol{$\triangleleft$}
\newtheorem{theorem}[dummy]{Theorem}
\newtheorem{lemma}[dummy1]{Lemma}
\newtheorem{definition}[dummy2]{Definition}
\newtheorem{proposition}[dummy3]{Proposition}

\theorembodyfont{\normalfont}
\newtheorem{example}[dummy4]{Example}
\newtheorem{remark}[dummy5]{Remark}

\theoremstyle{nonumberplain}
\theoremheaderfont{\itshape}
\theorembodyfont{\normalfont}

\theoremseparator{.}
\theoremsymbol{\ensuremath{\dashv}}
\newtheorem{proof}[dummy6]{Proof}

\qedsymbol{\ensuremath{\dashv}}

\newcommand{\dab}{DAB\xspace}

\newcommand{\ua}{\ensuremath{\underline a}}

\newcommand{\ue}{\ensuremath{\underline e}}
\newcommand{\ui}{\ensuremath{\underline i}}
\newcommand{\uu}{\ensuremath{\underline u}}
\newcommand{\uv}{\ensuremath{\underline v}}
\newcommand{\ux}{\ensuremath{\underline x}}
\newcommand{\uy}{\ensuremath{\underline y}}
\newcommand{\uz}{\ensuremath{\underline z}}

\newcommand{\uS}{\ensuremath{\underline S}}

\newcommand{\PSPACE}{\textsc{Pspace}\xspace}

\newcommand{\true}{\constant{true}\xspace}
\newcommand{\false}{\constant{false}\xspace}

\newcommand{\cM}{\ensuremath \mathcal M}

\newcommand{\cS}{\ensuremath \mathcal S}

\newcommand{\safe}{\texttt{SAFE}\xspace}
\newcommand{\unsafe}{\texttt{UNSAFE}\xspace}
\newcommand{\mcmt}{\textsc{mcmt}\xspace}

\renewcommand{\int}{\ensuremath {\mathcal I}}

\newcommand{\domain}[1]{\mathit{dom}(#1)}

\newcommand{\sas}{SAS\xspace}

\newcommand{\ras}{RAS\xspace}

\newcommand{\relname}[1]{\ensuremath{\mathit{#1}}\xspace}

\newcommand{\constant}[1]{\texttt{#1}}
\newcommand{\sorts}[1]{#1}
\newcommand{\sort}[1]{\sorts{\mathsf{#1}}}
\newcommand{\functs}[1]{#1_{\mathit{fun}}}
\newcommand{\vals}[1]{#1_{\mathit{val}}}
\newcommand{\ids}[1]{#1_{\mathit{ids}}}
\newcommand{\ext}[1]{#1_{\mathit{ext}}}

\newcommand{\nullv}{\texttt{undef}}

\newcommand{\lentry}[2]{\ensuremath{#1\,{:}\,#2}}

\definecolor{deepblue}{HTML}{0C3B80}
\definecolor{deepgreen}{HTML}{2EA601}
\definecolor{lightOrange}{HTML}{FFA03C}
\definecolor{darkOrange}{HTML}{F1800A}
\definecolor{lightBlue}{HTML}{0174CD}
\definecolor{greenF}{HTML}{2CBB5C}
\definecolor{cyan}{HTML}{86A6D5}

\tikzstyle{sortnode} = [
  rectangle,
  rounded corners=5pt,
  minimum width=18mm,
  minimum height=6mm,
  very thick,
]

\tikzstyle{functnode} = [
  rectangle,
  minimum width=1cm,
  minimum height=5mm,
]

\tikzstyle{idnode} = [
  sortnode,
  draw=deepblue,
  fill=cyan!50,
  text=deepblue,  
]

\tikzstyle{valnode} = [
  sortnode,
  densely dashed,
  draw=violet,
  fill=magenta!50,
  text=violet,
]

\tikzstyle{f} = [
  thick,
]

\tikzstyle{fd} = [
  f,
  -angle 90,
]

\tikzstyle{relation}=[rectangle split, rectangle split parts=#1, rectangle split part align=base, draw, anchor=center, align=center, text height=3mm, font=\bfseries, text centered]

\newcommand{\typedvar}[2]{#1{:}#2}

\newcommand{\eff}{\mathit{Eff}}

\newcommand{\self}{\cvar{self}}
\newcommand{\PI}{\mathcal{PI}}
\newcommand{\SET}{\xspace\mathtt{SET}\xspace}
\newcommand{\INSERT}{\mathtt{INSERT}\xspace}
\newcommand{\INTO}{\mathtt{INTO}\xspace}
\newcommand{\FROM}{\mathtt{FROM}\xspace}

\newcommand{\MOVE}{\mathtt{DEL}\xspace}
\newcommand{\TO}{\mathtt{TO}\xspace}
\newcommand{\SETTING}{\mathtt{AND~SET}\xspace}
\newcommand{\UPDATE}{\mathtt{UPDATE}\xspace}
\newcommand{\IF}{\xspace\mathtt{IF}\xspace}
\newcommand{\THEN}{\xspace\mathtt{THEN}\xspace}
\newcommand{\ELSE}{\xspace\mathtt{ELSE}\xspace}
\newcommand{\nvar}[1]{\mathit{?}#1\xspace}

\tikzstyle{sortnode} = [
  rectangle,
  rounded corners=5pt,
  minimum width=18mm,
  minimum height=6mm,
  very thick,
]

\tikzstyle{functnode} = [
  rectangle,
  minimum width=1cm,
  minimum height=5mm,
]

\tikzstyle{idnode} = [
  sortnode,
  draw=deepblue,
  fill=cyan!50,
  text=deepblue,  
]

\tikzstyle{artnode} = [
  sortnode,
  draw=deepblue,
  fill=yellow!50,
  text=deepblue,  
]

\tikzstyle{valnode} = [
  sortnode,
  densely dashed,
  draw=violet,
  fill=magenta!50,
  text=violet,
]

\tikzstyle{f} = [
  thick,
]

\tikzstyle{fd} = [
  f,
  -angle 90,
]

\tikzstyle{relation}=[rectangle split, rectangle split parts=#1, rectangle split part align=base, draw, anchor=center, align=center, text height=3mm, font=\bfseries, text centered]

\tikzstyle{task} = [minimum height=1cm,minimum width = 1.4cm,rectangle,fill=white,thick,draw,rounded corners,align=center, every task, font=\footnotesize]
\tikzstyle{sequence} = [->,>=triangle 45,every sequence,thick]
\tikzstyle{guard} = [rectangle,draw=none,fill=white,inner sep=.5mm,minimum height=.8mm,minimum width=.8mm]
\tikzstyle{lbl} = [text width=4cm]
\newcommand{\tname}[1]{\textsf{#1}} 
\newcommand{\ename}[1]{\textsl{#1}}

\begin{document}


\maketitle


\begin{abstract}
We propose DAB -- a data-aware extension of the BPMN de-facto standard with the ability of operating over case and persistent data (partitioned into a read-only catalog and a read-write repository), and that balances between expressiveness and the possibility of supporting parameterized verification of safety properties on top of it. In particular, we take inspiration from the literature on verification of artifact systems, and consider verification problems where safety properties are checked irrespectively of the content of the read-only catalog, possibly considering an unbounded number of active cases and tuples in the catalog and repository. Such problems are tackled using fully implemented array-based backward reachability techniques belonging to the well-established tradition of SMT model checking. We also identify relevant classes of DABs for which the backward reachability procedure implemented in the MCMT array-based model checker is sound and complete, and then further strengthen such classes to ensure termination.
\end{abstract}

\section{Introduction}

In recent years, increasing attention has been given to multi-perspective models of business processes that strive to capture the interplay between the process and data dimensions~\cite{Rich10,Reic12}. The corresponding development of formal models and foundational results on their verification has flourished, but mainly focusing on data-centric approaches \cite{Vian09,CaDM13} that are quite different in nature from conventional notations. In parallel, conventional approaches such as the de-facto BPMN standard have been extended with various forms of data, though with a main focus on conceptual modeling and enactment \cite{MPFW13,DOET17,COWZ18}, without considering formal verification. More on the formal side, many approaches within this line considering Petri nets as the main control-flow backbone for capturing the process, and enriching them with data locally carried by tokens \cite{RVFE11,Las16,MonR16}, case data with different datatypes \cite{DeFM18}, and/or persistent relational data manipulated with the full power of FOL/SQL \cite{DDG16,MonR17}. While this latter family of approaches qualify well to capture BPMN enriched with persistent data (such as \cite{MPFW13,COWZ18}), they place two assumptions on verification: first, that the process is studied by considering a fully-specified, initial instance of the underlying database; second, that only boundedly many tuples can be stored in the database. Such approaches have not yet been applied to formalize BPMN with data, and do not lend themselves to capture parameterized verification problems that can ascertain the correctness of the process without imposing the aforementioned limitations. 

In this work, we attack this open challenge and propose a data-aware extension of BPMN, called \emph{data-aware BPMN} (\dab), equipped with the ability of querying and operating over case and persistent data (partitioned into a read-only catalog and a read-write repository). The approach balances expressiveness with the possibility of supporting parameterized verification of safety properties. In particular, we take inspiration from the literature on the verification of artifact systems \cite{Vian09}, and in particular the frameworks in \cite{verifas,CGGMR19},  considering verification problems where safety properties are checked irrespectively of the content of the read-only catalog, and possibly considering an unbounded number of active cases and tuples in the catalog and repository.

We study this problem by establishing a bridge between our approach to business process modeling, and the line of research where techniques based on Satisfiability Modulo Theories (SMT) are employed to attack the verification of infinite-state \emph{array-based systems} -- originally introduced in~\cite{ijcar08,lmcs} to handle the
verification of  distributed systems (parameterized on the number of interacting 
processes). In particular, we rely on novel foundational results on the verification of artifact-centric systems via array-based systems \cite{CGGMR18,CGGMR19}, which have been fully implemented in the state-of-the-art \textsc{mcmt} SMT symbolic model checker \cite{mcmt}.

By exploiting this formal basis, we identify two relevant classes of DABs for which the backward reachability procedure implemented in the MCMT array-based model checker is sound and complete. This guarantees that if the procedure terminates, it produces a correct judgement.  We then introduce further conditions that, by carefully controlling the interplay between the process and data components, guarantee the termination of the procedure, in turn witnessing decidability. Such conditions are expressed as syntactic restrictions over the \dab under study, thus providing a concrete, BPMN-grounded counterpart of the conditions imposed in \cite{verifas,CGGMR18,CGGMR19} towards decidability. 

These termination results obtained by translating DABs into the array-based artifact systems studied in \cite{CGGMR18,CGGMR19}. The translation can then be straightforwardly implemented making it possible to effectively verify DABs using \textsc{mcmt}.

This article builds upon \cite{Arxiv-BPM}, extending it in two respects. On the one hand, while \cite{Arxiv-BPM} focuses on the verification of DABs considering a single, running case, we consider here the possibility of (unboundedly many) cases running concurrently. On the other hand, we provide full proofs of the technical results, including those from  \cite{Arxiv-BPM} and those specifically introduced in this extended version.


\newcommand{\types}{\mathcal{S}}
\newcommand{\atype}{S}

\newcommand{\cat}{\mathit{Cat}}
\newcommand{\caseid}{\mathit{CType}}
\newcommand{\repo}{\mathit{Repo}}
\newcommand{\cvars}{\mathit{X}}

\newcommand{\vtypes}{\types_{v}}
\newcommand{\idtypes}{\types_{id}}
\newcommand{\datadom}{\mathbb{D}}
\newcommand{\vdatadom}{\datadom_v}
\newcommand{\iddatadom}{\datadom_{id}}

\newcommand{\cvar}[1]{\mathbf{#1}}
\renewcommand{\nvar}[1]{\mathit{#1}}

\newcommand{\dotcomp}[1]{\mathsf{#1}}
\newcommand{\dotexpr}[2]{#1.\dotcomp{#2}}

\newcommand{\reln}[1]{\dotexpr{\relname{#1}}{name}}
\newcommand{\relid}[1]{\dotexpr{\relname{#1}}{id}}
\newcommand{\relattrs}[1]{\dotexpr{\relname{#1}}{attrs}}

\newcommand{\dcat}[1]{\dotexpr{#1}{cat}}
\newcommand{\drepo}[1]{\dotexpr{#1}{repo}}
\newcommand{\dcvars}[1]{\dotexpr{#1}{cvars}}
\newcommand{\dctype}[1]{\dotexpr{#1}{ctype}}

\newcommand{\freevars}[1]{\mathit{free}(#1)}
\newcommand{\getdatavars}[1]{\mathit{normvars}(#1)}
\newcommand{\getcasevars}[1]{\mathit{casevars}(#1)}

\newcommand{\tpre}{G}
\newcommand{\tpost}{E}

\newcommand{\getpre}[1]{\dotexpr{#1}{pre}}
\newcommand{\getpost}[1]{\dotexpr{#1}{eff}}

\newcommand{\updatespec}[1]{\texttt{#1}}

\newcommand{\dhr}{\D^h}

\newcommand{\bldist}{-1mm}
\newcommand{\outfdist}{4mm}
\tikzstyle{outflow} = [sequence,->,densely dotted]
\tikzstyle{smalltask} = [rectangle,draw,rounded corners=5pt,minimum height=2.5em,minimum width=3em]
\tikzstyle{block} = [task,densely dotted]
\tikzstyle{smallblock} = [smalltask,densely dotted]
\tikzstyle{legend} = [font=\footnotesize]

\tikzstyle{state} = [
  rectangle,
  draw,
  rounded corners=10pt,
  minimum width=15mm,
  minimum height=7mm]

\newcommand{\sidle}{\constant{idle}}
\newcommand{\senabled}{\constant{enabled}}
\newcommand{\sactive}{\constant{active}}
\newcommand{\scompleted}{\constant{compl}}
\newcommand{\scomplerror}{\constant{error}}

\section{Data-aware BPMN}



We start by describing our formal model of data-aware BPMN processes (\dab{s}). We focus here on private, single-pool processes. Incoming messages are therefore handled as pure nondeterministic events. The model combines a wide range of (block-structured) BPMN control-flow constructs with task, event-reaction, and condition logic that inspect and modify persistent as well as case data. The combination achieves a balanced trade off between the expressive power of the resulting integrated model, and the possibility of carrying out sophisticated forms of parameterized verification, which will be tackled in Section~\ref{sec:verification}. When going through the modeling features of \dab, it is then important to remember that if something is not supported, it is because it would hamper soundness and completeness of SMT-based (parameterized) verification.

First, some preliminary notation. We consider a set $\types = \vtypes \uplus \idtypes$ of (semantic) \emph{types}, consisting of \emph{primitive types} $\vtypes$ accounting for data objects, and \emph{id types} $\idtypes$ accounting for identifiers. We assume that each type $\atype \in \types$ comes with a (possibly infinite) domain $\datadom_\atype$, a special constant $\nullv_\atype \in \datadom_\atype$ to denote an undefined value in that domain, and a type-wise equality operator $=_\atype$. We omit the type and simply write $\nullv$ and $=$ when clear from the context. We do not consider here additional type-specific predicates (such as comparison and arithmetic operators for numerical primitive types); these will be added in future work (cf.~Section~\ref{sec:conclusion} for a discussion on this). In the following, we simply use \emph{typed} as a shortcut for $\types$-\emph{typed}. We also  denote by $\datadom$ the overall domain of objects and identifiers (i.e., the union of all domains in $\types$). 
We consider a countably infinite set $\V$ of typed variables. Given a variable or object $x$, we may explicitly indicate that $x$ has type $\atype$ by writing $x:\atype$. We omit types whenever clear from the context, or irrelevant. We compactly indicate a possibly empty tuple $\tup{x_1,\ldots,x_n}$ of variables as $\vec{x}$, and with slight abuse of notation, we write $\vec{x} \subseteq \vec{y}$ if all variables in $\vec{x}$ also appear in $\vec{y}$.

\subsection{The Data Schema}
\label{sec:data-schema}
Consistently with the BPMN standard, we consider two main forms of data: \emph{case data}\footnote{These are called \emph{data objects} in BPMN, but we prefer to use the term \emph{case data} to avoid name clashes with the formal notions.}, instantiated and manipulated on a per-case basis; \emph{persistent data} (cf.~data store references in BPMN), accounting for global data that are accessed by all cases. For simplicity, case data are defined at the whole process level, and are directly visible by all tasks and subprocesses (without requiring the specification of input-output bindings and the like).

To account for persistent data, we consider relational databases. We describe relation schemas by using the \emph{named perspective}, i.e., by assigning a dedicated typed attribute to each component (i.e., column) of a relation schema. Also for an attribute, we use the notation $a:\atype$ to explicitly indicate its type. 
\begin{definition} 
\label{def:relation-schema}
A \emph{relation schema} is a pair $R=\tup{N,A}$, where:
\begin{inparaenum}[\it (i)]
\item $N = \reln{R}$ is the relation \emph{name};
\item $A = \relattrs{R}$ is a nonempty tuple of attributes.  
\end{inparaenum}
We call $|A|$ the \emph{arity} of $R$. 
\end{definition}
We assume that distinct relation schemas use distinct names, blurring the distinction between the two notions (i.e., we set $\reln{R} = R$). We also use the predicate notation $R(A)$ to represent a relation schema $\tup{R,A}$.

\inlinetitle{Data schema}
We start by defining the \emph{catalog}, i.e., a read-only, persistent storage of data that is not modified during the execution of the process. 
\begin{definition}
\label{def:catalog}
A \emph{catalog} $\cat$ is a set of relation schemas satisfying the following requirements:
\begin{compactdesc}
  \item[(single-column primary key)] Every relation schema $R$ is such that the first attribute in $\relattrs{R}$ has type in $\idtypes$, and denotes the \emph{primary key} of the relation; we refer to such attribute using the dot notation $\relid{R}$.
  \item[(non-ambiguity of primary keys)] for every pair $R_1$ and $R_2$ of \emph{distinct} relation schemas in $\cat$, we have that the types of $\relid{R_1}$ and $\relid{R_2}$ are different.
  \item[(foreign keys)] for every relation schema $R \in \cat$ and non-id attribute $a \in \relattrs{R} \setminus \relid{R}$ with type $\atype \in \idtypes$, there exists a relation schema $R_2 \in \R$ such that the type of $\relid{R_2}$ is $\atype$; $a$ is hence a \emph{foreign key} referring to $R_2$.
\end{compactdesc}  
\end{definition}

We now define the data schema of a BPMN process, which combines a catalog with:
\begin{inparaenum}[\it (i)]
\item a persistent data \emph{repository}, consisting of updatable relation schemas possibly referring to the catalog;
\item a set of \emph{case variables}, constituting local data carried by each process case.  
\end{inparaenum}



\begin{definition}
\label{def:data-component}
A \emph{data schema} $\D$ is a tuple $\tup{\cat,\caseid,\repo,\cvars}$, where
\begin{inparaenum}[\it (i)]
\item $\cat = \dcat{\D}$ is a \emph{catalog},
\item $\caseid = \dctype{\D} \in \idtypes$ is a special \emph{case identifier type},   
\item $\repo = \drepo{\D}$ is a set of relation schemas called \emph{repository}, and
\item $\cvars = \dcvars{\D} \subset \V$ is a finite set of typed variables called \emph{case variables},
\end{inparaenum} 
such that:
\begin{compactitem}[$\bullet$]
\item $\caseid$ is disjoint from all identifier types used in $\cat$;
\item for every relation schema $R \in \repo$ and every attribute $a \in \relattrs{R}$ whose type is $S \in \idtypes$, there exists $R \in \cat \cup \set{\caseid}$ such that the type of $\relid{R}$ is $S$;
\item for every case variable $\cvar{x} \in \cvars$ whose type is $S \in \idtypes$, there exists $R \in \cat \cup \set{\caseid}$ such that the type of $\relid{R}$ is $S$;
\item $\D$ contains an special case variable $\self{:}\caseid$ that is never modified, and that keeps track, for a case, of the corresponding case identifier.
\end{compactitem}
\end{definition}
We use bold-face to distinguish a case variable $\cvar{x}$ from a ``normal" variable $x$. 
It is worth noting that relation schemas in the repository are not equipped with an explicit primary key, and thus they cannot reference each other, but may contain foreign keys pointing to the catalog or the case identifiers. \emph{This is essential towards soundness and completeness of SMT-based verification of \dab{s}}. It will be clear how tuples can be inserted and removed from the repository once we will introduce updates.

At runtime, a \emph{data snapshot} of a data schema consists of three components:
\begin{compactitem}[$\bullet$]
\item An immutable \emph{catalog instance}, i.e., a fixed set of tuples for each relation schema contained therein, so that the primary and foreign keys are satisfied. 
\item A \emph{case map} whose keys are the identifiers of active or completed cases (i.e., elements of the case identifier type), and whose values are assignments of the case variables to corresponding values (satisfying the foreign keys when pointing to identifiers in the catalog). Each entry then indicates, for a given case, which are the current values for the case variables of that case. 
\item A \emph{repository instance}, i.e., a set of tuples for for each relation schema contained therein, so that the foreign key constraints pointing to the catalog or the case map keys are satisfied. Each tuple is associated to a distinct primary key that is not explicitly accessible.
\end{compactitem}

\begin{example}
  \label{ex:data-schema}
  We consider a simplified example of a job hiring process in a company. We describe here the data schema $\dhr$ used to store data about job hirings and their corresponding applications. The catalog $\dcat{\dhr}$ consists of the following relation schemas:
\begin{compactitem}[$\bullet$]
\item $\relname{JobCategory}(Jcid{:}\sort{jobcatID})$ contains the different job categories available in the company (e.g., programmer, analyst, and the like) - we just store here the identifiers of such categories;
\item $\relname{User}(Uid{:}\sort{userID},Name{:}\sort{StringName},Age{:}\sort{NumAge})$ stores data about users registered to the company website, and who are potentially interested in job positions offered by the company.
\end{compactitem}
Each case of the process is about a job. Jobs are identified by the type $\sort{jobId}$.

To manage key information about the applications submitted for the various job hirings, including data on users, the score they receive after having been interviewed and their eligibility, the company employs the repository $\drepo{\D}$ that consists of one relation schema 
\[
\relname{Application}(
\begin{array}[t]{@{}l@{}}
Jid{:}\sort{jobId},Jcid{:}\sort{jobcatID},Uid{:}\sort{userID},Name{:}\sort{StringName},Age{:}\sort{NumAge},
\\
Score{:}\sort{NumScore},Eligible{:}\sort{Bool})
\end{array}
\] Notice that $\sort{NumScore}$ is a finite-domain type containing $100$ values in the range
      $[\constant{1},\constant{100}]$, and it is used to assign an overall score to each candidate application. For readability, we use the usual predicates $<$, $>$,
  and $=$ to compare variables of type $\sort{NumScore}$: this is syntactic sugar and does not require to introduce rigid predicates in our framework. 

Since each posted job is created using a dedicated portal, its corresponding data do not have to be stored persistently and thus can be maintained just for a given case. At the same time, some specific values have to be moved from a specific case to the repository and vice-versa. This is done by resorting to the following case variables $\dcvars{\D}$:
\begin{inparaenum}[\it (i)]
\item $\cvar{jcid}:\sort{jobcatID}$ references a job type from the catalog, matching the type of job associated to the case;
\item $\cvar{uid}:\sort{userID}$, $\cvar{name}:\sort{StringName}$ and $\cvar{age}:\sort{NumAge}$ respectively 
reference the identifier, name, and age of a user who is applying for the job associated to the case;
\item $\cvar{result}:\sort{Bool}$ indicates whether the user identified by $\cvar{uid}$ is eligible for winning the position or not;
\item $\cvar{result}:\sort{Bool}$ indicates whether the user identified by $\cvar{uid}$ qualifies for directly getting the job (without the need of carrying out a comparative evaluation of all applicants); 
\item $\cvar{winner}:\sort{userID}$ contains the identifier of the applicant winning the position;
\item $\cvar{tPassed}:\sort{StringDate}$ contains special strings symbolically indicating the current temporal phase of the case in relation with its creation time.
\end{inparaenum}
The last variable is not essential for the progression of job hirings through the process, but it is useful to formulate verification properties. 
\end{example}

\inlinetitle{Querying the data schema}
To inspect the data contained in a snapshot, we need suitable query languages operating over the data schema of that snapshot. In the following, we assume that queries are well-typed, i.e., sorts of their elements are duly matched (this can be easily checked by scanning the query).
We start by considering boolean \emph{conditions} over (case) variables. These conditions will be attached to choice points in the process.

\begin{definition}
  \label{def:condition}
  A \emph{condition} is a formula of the form $\varphi ::= (x=y) \mid \neg \varphi \mid \varphi_1 \land \varphi_2$, where $x$ and $y$ are variables from $\V$ or constant objects from $\datadom$. If in $\varphi$ negation is restricted to be only in front of atoms, $\varphi$ is called a \emph{cubical condition}.
\end{definition}

We make use of the standard abbreviation $\varphi_1 \lor \varphi_2 = \neg (\neg \varphi_1 \land \neg \varphi_2)$. 

We now extend conditions to also access the data stored in the catalog and repository, and to ask for data objects subject to constraints. We consider the well-known language of unions of conjunctive queries with atomic negation, which correspond to unions of select-project-join SQL queries with table filters. 
%

\begin{definition}
A \emph{conjunctive query with filters} over a data component $\D$ is a formula of the form
  $Q ::= \varphi \mid R(x_1,\ldots,x_n) \mid \neg R(x_1,\ldots,x_n) \mid Q_1 \land Q_2$, where $\varphi$ is a cubical condition, $R\in  \dcat{\D} \cup \drepo{\D}$ is a relation schema of arity $n$, and $x_1,\ldots,x_n$ are variables from $\V$ (including $\dcvars{\D}$) or constant objects from $\datadom$. We denote by $\freevars{Q}$ the set of variables occurring in $Q$ that are \emph{not} case variables in $\dcvars{\D}$.  
\end{definition}

\begin{definition}
\label{def:guard}
  A \emph{guard} $G$ over a data component $\D$ is an expression of the form $q(\vec{x}) \leftarrow \bigvee_{i=1}^n Q_i$, where:
  \begin{inparaenum}[\it (i)]
 \item $q(\vec{x})$ is the \emph{head} of the guard with \emph{answer variables} $\vec{x}$;
  \item each $Q_i$ is a conjunctive query with filters over $\D$; 
  \item for each $i \in \set{1,\ldots,n}$, $\vec{x} \subseteq \freevars{Q_i}$.
  \end{inparaenum} 
  We denote by $\getcasevars{G} \subseteq \dcvars{\D}$ the set of case variables used in $G$, and by $\getdatavars{G} = \bigcup_{i \in \set{1,\ldots,n}} \freevars{Q_i}$ the other variables used in $G$. 
\end{definition}

\begin{definition}
  A \emph{guard} $G$ over a data component $\D$ is \emph{repo-free} if none of its atoms queries a relation schema from $\drepo{\D}$.
\end{definition}

Notice that \emph{going beyond this guard query language} (e.g., by introducing universal quantification) \emph{would hamper the soundness and completeness of SMT-based verification over the resulting \dab{s}}. We will come back to this important aspect in the conclusion. 

As anticipated before, this language can be seen as a standard query language to retrieve data from a snapshot, but also as a mechanism to identify the allowed combinations of data objects that can be injected into the process from the external environment. For example, considering a case variable $\cvar{x}$ of type $\sort{string}$, a simple guard $\relname{Input}(y{:}\sort{string},z{:}\sort{string}) \rightarrow y \neq \cvar{x} \land y \neq z$ returns all pairs of strings that are different from each other, and so that the second string is different from that stored in the case variable $\cvar{x}$. Picking an answer in this (infinite) set of pairs can be consequently seen as a (constrained) user decision on which values to input for $y$ and $z$. We  elaborate more on this in Section~\ref{sec:update-logic}.

%

\subsection{Tasks, Events, and Impact on Data}
\label{sec:update-logic}
We now formalize how the process can access and update the data component when executing a task or reacting to the trigger of an external event. 

\inlinetitle{The update logic}
We start by discussing how data maintained in a snapshot can be subject to change while executing the process.
\begin{definition}
\label{def:update-spec}
Given a data schema $\D$, an \emph{update specification} $\alpha$ is a pair $\tup{\tpre,\tpost}$, where:
\begin{inparaenum}[\it (i)]
\item $\tpre = \getpre{\alpha}$ is a guard over $\D$ of the form $q(\vec{x}) \leftarrow Q$, called \emph{precondition}; 
\item $\tpost= \getpost{\alpha}$ is an \emph{effect rule} that changes the content of case variables or that of the repository, as described next.
\end{inparaenum} 
Each effect rule has one of the following forms:
\begin{compactdesc}
\item[(Insert\&Set)] $\INSERT~\vec{u}~\INTO~R~\SETTING~\cvar{x}_1 = v_1, \ldots, \cvar{x}_n = v_n$, where: 
\begin{inparaenum}[\it (i)] 
\item  $\vec{u}, \vec{v}$ are variables in $\vec{x}$ or constant objects from $\datadom$; 
\item $\vec{\cvar{x}} \in \dcvars{\D}\setminus\set{\self}$ are distinct case variables different from $\self$; 
\item $R$ is a relation schema from $\drepo{\D}$ whose arity (and types) match $\vec{u}$. 
\end{inparaenum}
Either the $\INSERT$ or $\SETTING$ parts may be omitted, obtaining a pure (repository) \textnormal{\textbf{Insert rule}} or (case variable) \textnormal{\textbf{Set rule}}.
\item[(Delete\&Set)] $\MOVE~\vec{u}~\FROM~R~\SETTING~\cvar{x}_1 = v_1, \ldots, \cvar{x}_n = v_n$, where:
 \begin{inparaenum}[\it (i)] 
\item  $\vec{u},\vec{v}$ are variables in $\vec{x}$ or constant objects from $\datadom$; ;
\item  $\vec{\cvar{x}} \in \dcvars{\D}\setminus\set{\self}$; 
\item $R$ is a relation schema from $\drepo{\D}$ whose arity (and types) match $\vec{u}$. 
\end{inparaenum}
As in the previous rule type, the $\SETTING$ part may be omitted, obtaining a pure (repository) \textnormal{\textbf{Delete rule}}.
\item[(Conditional update)] $\UPDATE~R(\vec{v})~\IF~\psi(\vec{u},\vec{v})~\THEN~\eta_1~\ELSE~\eta_2$, where:
 \begin{inparaenum}[\it (i)] 
\item $\vec{u}$ is a tuple containing variables in $\vec{x}$ or constant objects from $\datadom$; ;
\item $\psi$ is a repo-free guard (called \emph{filter})
\item $\vec{v}$ is a tuple of new variables, i.e., such that $\vec{v} \cap (\vec{u} \cup \dcvars{\D}) = \emptyset$;
\item $\eta_i$ is either an atomic formula of the form $R(\vec{u}')$ with $\vec{u}'$ a tuple of elements from $\vec{x} \cup \datadom \cup \vec{v}$, or a nested $\IF\ldots\THEN\ldots\ELSE$.
\end{inparaenum}
\end{compactdesc}
\end{definition}
As stated in the definition, $\self$ is never explicitly set by any of the effect rules.

We now comment on the semantics of update specifications. An update specification $\alpha$ is executable in a given data snapshot if there is at least one answer to the precondition $\getpre{\alpha}$ in that snapshot. If this is the case, then the process executor(s) can nondeterministically decide which answer to pick so as to \emph{bind} the answer variables of $\getpre{\alpha}$ to corresponding data objects in $\datadom$. This confirms the interpretation discussed in Section~\ref{sec:data-schema} for which  the answer variables of $\getpre{\alpha}$ can be seen as \emph{constrained user inputs} in case multiple bindings are available. 

Once a specific binding for the answer variables is selected, the corresponding effect rule $\getpost{\alpha}$, instantiated using that binding, is issued. How this affects the current data snapshot depends on which effect rule is adopted.

If  $\getpost{\alpha}$ is an insert\&set rule, the binding is used to \emph{simultaneously}
insert a tuple in one of the repository relations, and update some of the case variables -- with the implicit assumption that those not explicitly mentioned in the $\SET$ part  maintain their current values. 
Since repository relations do not have an explicit primary key, two possible semantics can be attached to the insertion of a tuple $\vec{u}$ in the instance of a repository relation $R$:
\begin{compactdesc}
\item[(multiset insertion)] Upon insertion, $\vec{u}$ gets implicitly assigned to a fresh primary key. The insertion then always results in the genuine addition of the tuple to the current instance of $R$, even in the case where the tuple already exists there.
\item[(set insertion)] In this case, $R$ comes not only with its implicit primary key, but also with an additional, genuine key constraint defined over a subset $K \subseteq \relattrs{R}$ of its attributes. Upon insertion, if there already exists a tuple in the current instance of $R$ that agrees with $\vec{u}$ on $K$, then that tuple is \emph{updated} according to $\vec{u}$. If no such tuple exists, then as in the previous case $\vec{u}$ gets implicitly assigned to a fresh primary key, and inserted into the current instance of $R$. By default, if no explicit key is defined over $R$, then the entire set of attributes $\relattrs{R}$ is considered as a key, consequently enforcing a \emph{set semantics} for insertion.
\end{compactdesc}

\begin{example}
\label{ex:updates-1}
  We continue the job hiring example, by considering two update specifications of type insert\&set. When a new case is created, the first update is about indicating what is the category of job associated to the case. This is done through the update specification $\updatespec{InsJobCat}$, where:
  \begin{compactitem}
  \item $\getpre{\updatespec{InsJobCat}} \triangleq \relname{GetJobType}(jt) \leftarrow \relname{JobCategory}(jt)$ selects a job category from the corresponding catalog relation;  
  \item $\getpost{\updatespec{InsJobCat}} \triangleq \SET~\cvar{jcid} = jt$ assigns the selected job category $jt$ to the case variable $\cvar{jcid}$.
  \end{compactitem}
  When the case receives an application for its associated job, the user-related case variables are filled with the data of the user submitting the application - picked from the corresponding $\mathit{User}$ catalog relation. This is done via the update specification $\updatespec{InsUser}$, where:
  $$
    \begin{array}{@{}rl@{}}
      \getpre{\updatespec{InsUser}} \triangleq & 
        \relname{GetUser}(u,n,a) 
        \leftarrow 
        \relname{User}(u,n,a)\\
      \getpost{\updatespec{InsUser}} \triangleq &
         \SET~\cvar{uid} = u, \cvar{name} = n, \cvar{age} = a\\
    \end{array}
  $$
  A different usage of precondition, resembling a pure external choice, is the update specification $\updatespec{CheckQual}$ to handle a quick evaluation of the candidate and chec whether she has such a high profile qualifying her to directly get an offer:
  $$
    \begin{array}{@{}rl@{}}
      \getpre{\updatespec{CheckQual}} \triangleq & 
        \relname{IsQualified}(q:\sort{Bool}) 
        \leftarrow 
        \true \\
      \getpost{\updatespec{CheckQual}} \triangleq &
         \SET~\cvar{qualif} = q\\
    \end{array}
  $$
  As an example of insertion rule, we consider the situation where the candidate whose data are currently stored in the case variables has not been directly judged as qualified. She is consequently subject to a more fine-grained evaluation of her application, resulting in a score that is then registered in the repository (together with the applicant data). This is done via the $\updatespec{Reg<eval}$ specification:
  $$
    \begin{array}{@{}rl@{}}
      \getpre{\updatespec{EvalApp}} \triangleq & 
        \relname{GetScore}(s:\sort{NumScore}) 
        \leftarrow 
        1 \leq s \land s \leq 100 \\
      \getpost{\updatespec{EvalApp}} \triangleq &
       \INSERT~\tup{\self,\cvar{jcid},\cvar{uid},\cvar{name},\cvar{age},s,\nullv}~\INTO~\relname{Application}\\
    \end{array}
  $$
  Here, the insertion uses the applicant data currently stored in the corresponding case variables, the selected score, and $\nullv$ eligibility (which is then assessed in a consequent step of the process). These objects are correlated to the case identifier, so as to keep track of the relationship between the application and the job to which the application has been submitted. This is essential, as the same user may apply for different jobs. Notice that, by adopting the \emph{multiset insertion semantics}, the same user may even apply multiple times for the same job. With a \emph{set insertion semantics}, one could instead ensure that each user can apply at most once to the same job, by indicating that the first two components of $\relname{Application}$ form a key.
  \end{example}

If $\getpost{\alpha}$ is a delete\&set rule, then the executability of the update is subject to the fact that the tuple $\vec{u}$ selected by the binding and to be removed from $R$, is actually present in the current instance of $R$. If so, the binding is used to \emph{simultaneously} delete $\vec{u}$ from $R$ and update some of the case variables -- with the implicit assumption that those not explicitly mentioned in the $\SET$ part maintain their current values. 

Finally, a conditional update rule applies, tuple by tuple, a bulk operation over the content of $R$. For each tuple in $R$, if it passes the filter associated to the rule, then the tuple is updated according to the $\THEN$ part, whereas if the filter evaluates to false, the tuple is updated according to the $\ELSE$ part.

\begin{example}
\label{ex:updates-2}
  Continuing with our running example, we now consider the update specification $\updatespec{MarkE}$ handling the situation where no candidate has been directly considered as qualified, and so the eligibility of all received (and evaluated) applications has to be assessed. Here we consider that each application is eligible if and only if its evaluation resulted in a score greater than $80$. Technically, $\getpre{\updatespec{MarkE}}$ is a true precondition, and:
  $$
    \getpost{\updatespec{MarkE}} \triangleq \begin{array}[t]{@{}l@{}}
       \UPDATE~\relname{Application}(j,jc,u,n,a,s,e)\\
       \IF~j=\self \land s > 80~\THEN~\relname{Application}(j,jc,u,n,a,s,\true)\\
       \ELSE~\IF~j=\self \land s \leq 80~\THEN~\relname{Application}(j,jc,u,n,a,s,\false)\\
       \phantom{\ELSE~}\ELSE~\relname{Application}(j,jc,u,n,a,s,e)\\
    \end{array}
  $$  
  The update logic realized by $\getpost{\updatespec{MarkE}}$ is the following:
  \begin{inparaenum}[\it (i)]
  \item applications sent for the considered job and with a score $>80$ are marked as eligible;
  \item other applications sent for the considered job are marked as not eligible;
  \item applications sent for other jobs are left unaltered. 
  \end{inparaenum}
  
  If there is at least one eligible candidate, she can be selected as a winner using the $\updatespec{SelWinner}$ update specification, which deletes the selected winner tuple from $\relname{Application}$, and transfers its content to the corresponding case variables (also ensuring that the $\cvar{winner}$ case variable is set to the applicant id). Technically:
  $$
    \begin{array}{@{}rl@{}}
      \getpre{\updatespec{SelWinner}} \triangleq & 
        \relname{GetWinner}(j,jc,u,n,a,s,e) 
        \leftarrow 
          \begin{array}[t]{@{}l@{}}
            \relname{Application}(j,jc,u,n,a,s,e)\\
            \land j = \self \land e = \true\\
          \end{array}\\
      \getpost{\updatespec{SelWinner}} \triangleq &
       \MOVE~\tup{j,jc,u,n,a,s,e}~\FROM~\relname{Application}\\
       &\SETTING~\cvar{jcid}= jc, \cvar{uid} = u,\cvar{name} = n,\\
       & \cvar{age} = a,\cvar{winner}=jc, \cvar{result}=e
    \end{array}
  $$
  Deleting the tuple is useful in the situation where the selected winner may refuse the job, and consequently should not be considered again if a new winner selection is carried out. To keep such tuple in the repository, one would just need to remove the $\MOVE$ part from $\getpost{\updatespec{EvalApp}}$.
  \end{example}

\inlinetitle{The task/event logic}
We now substantiate how the update logic is used to specify the task/event logic within a \dab process. The first important observation, which does not relate to our specific design choice for the update logic, but is inherently present whenever the process control flow is enriched with relational data, is that update effects manipulating the repository must be executed in an atomic, non-interruptible way. This is essential to ensure that insertions/deletions into/from the repository are applied on the same data snapshot where the precondition is checked. This cannot be guaranteed if the precondition and effect occur in different moments, as they may nondeterministically interleave with other update specifications potentially operating over the same portion of the repository.  
This is why in our approach we consider two types of task: \emph{atomic} and \emph{nonatomic}. This goes beyond the BPMN standard, where generic tasks are implicitly assumed to be nonatomic. 

Each atomic task/catching event is associated to a corresponding update specification. In the case of tasks, the specification precondition indicates under which circumstances the task can be enacted, and the specification effect how enacting the task impacts on the underlying data snapshot. In the case of events, the specification precondition constrains the data payload that comes with the event (possibly depending on the data snapshot, which is global and therefore accessible also from the perspective of an external event trigger), and the specification effect how reacting to a triggered event impacts on the underlying data snapshot. More concretely, this is realized according to the following lifecycle.

The task/event is initially $\sidle$, i.e., quiescent. When the progression of a case reaches an $\sidle$ task/event, such a task/event becomes $\senabled$. An $\senabled$ task/event may nondeterministically fire depending on the choice of the process executor(s). Upon firing, a binding satisfying the precondition of the update specification associated to the task/event is selected, consequently grounding and applying the corresponding effect. At the same time, the lifecycle moves from $\senabled$ to $\scompleted$. Finally, a $\scompleted$ task/event triggers the progression of its case depending on the process-control flow, simultaneously bringing the task/event back to the $\sidle$ state (which would then make it possible for the task to be executed again later on within the same case, if the process control-flow dictates so). 

The lifecycle of a nonatomic task diverges in two crucial respects. First of all, upon firing it moves from $\senabled$ to $\sactive$, and later on nondeterministically from $\sactive$ to $\scompleted$ (thus having a duration). The precondition of its update specification is checked and bound to one of the available answers when the task becomes $\sactive$, while the corresponding effect is applied when the task becomes $\scompleted$. Since these two transitions occur asynchronously, to avoid the aforementioned transactional issues we assume that the effect operates, in this context, only on case variables (and not on the repository).

\subsection{Process Schema}
A process schema consists of a block-structured BPMN diagram, enriched with conditions and update effects expressed over a given data schema, according to what described in the previous sections. As for the control flow, we consider a wide range of block-structured patterns compliant with the standard, taking inspiration and expanding those in \cite{MBMP18}. We focus on private BPMN processes, thereby handling incoming messages in a pure nondeterministic way. So we do for timer events, nondeterministically accounting for their expiration without entering into their metric temporal semantics. Focusing on block-structured components helps us in obtaining a direct,  execution semantics, and a consequent modular and clean translation of various BPMN constructs (including boundary events and exception handling). However, it is important to stress that our approach would seamlessly work also for non-structured processes where each case introduces boundedly many tokens. 

As usual, blocks are recursively decomposed into sub-blocks, the leaves being task or empty blocks. Depending on its type, a block may come with one or more nested blocks, and be associated with other elements, such as conditions, types of the involved events, and the like. We consider a wide range of blocks, covering basic, flow, and exception handling patterns. They are reported in Appendix~\ref{app:blocks}. Figure~\ref{fig:bpmn-hiring} gives an idea about what is covered by our approach.
 With these blocks at hand, we finally obtain the full definition of a \dab.

\begin{definition}
A \dab $\M$ is a pair $\tup{\D,\P}$ where $\D$ is a data schema, and $\P$ is a root \emph{process block} such that all conditions and update effects attached to $\P$ and its descendant blocks are expressed over $\D$.
\end{definition}

\begin{example}
  The full hiring job process is shown in Figured~\ref{fig:bpmn-hiring}, using the update effects described in Examples~\ref{ex:updates-1} and \ref{ex:updates-1}. Intuitively, the process works as follows. A new case for the process is created whenever a new job is posted. The case enters into a looping subprocess where it expects candidates to apply. Specifically, the case waits for an incoming application, or for an external message signalling that the hiring has to be stopped (e.g., because too much time has passed from the posting). Whenever an application is received, the CV of the candidate is evaluated, with two possible outcomes. The first outcome indicates that the candidate directly qualifies for the position, hence no further applications should be considered. In this case, the process continues by declaring the candidate as winner, and making an offer to her. The second outcome of the CV evaluation is instead that the candidate does not directly qualify. A more detailed evaluation is then carried out, assigning a score to the application and storing the outcome into the process repository, then waiting for additional applications to come. When the application management subprocess is stopped (which we model through an error so as to test various types of blocks in the experiments reported in Section~\ref{sec:mcmt}), the applications present in the working memory are all processed in parallel, declaring which candidates are eligible and which not depending on their scores. Among the eligible ones, a winner is then selected, making an offer to her. We implicitly assume here that at least one applicant is eligible. We can remove this assumption and also handle the case where no eligible applicant exists for a job, by simply introducing (and consequently using) a boolean case variable that, upon the application evaluation, is set to true if the obtained score makes the application eligible. \end{example}

As customary, each block has a lifecycle that, case by case, indicates the current state of the block, and how it can be evolved depending on the specific semantics of the block, and the evolution of its inner blocks. In Section~\ref{sec:update-logic} we have already characterized the lifecycle of tasks and catch events. For the other blocks, we continue to use the standard states $\sidle$, $\senabled$, $\sactive$ and $\scompleted$. We use the very same rules of execution described in the BPMN standard to regulate the progression of blocks through such states, taking advantage from the fact that, being the process block-structured, only one instance of a block can be enabled/active at a given time for a given case. For example, the lifecycle of a sequence block $\textsf{S}$ with nested blocks $\textsf{B}_1$ and $\textsf{B}_2$ can be described as follows (considering that the transitions of $\textsf{S}$ from $\sidle$ to $\senabled$ and from $\scompleted$ back to $\sidle$ are inductively regulated by its parent block):
\begin{inparaenum}[\it (i)]
\item if $\textsf{S}$ is $\senabled$, then it becomes $\sactive$, simultaneously inducing a transition  $\textsf{B}_1$ from $\sidle$ to $\senabled$;
\item if  $\textsf{B}_1$ is $\scompleted$, then it becomes $\sidle$, simultaneously inducing a transition of  $\textsf{B}_2$ from $\sidle$ to $\senabled$;
\item if  $\textsf{B}_2$ is $\scompleted$, then it becomes $\sidle$, simultaneously inducing $\textsf{S}$ to move from $\sactive$ to $\scompleted$.
\end{inparaenum}
The lifecycle of other block types can be defined analogously.

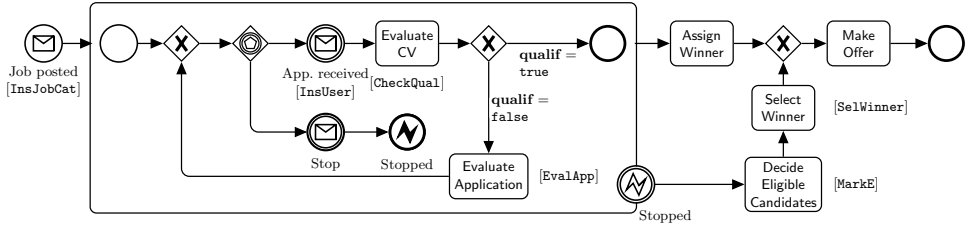
\begin{figure}[t!]
\hspace*{-1.3cm}
\resizebox{1.2\textwidth}{!}{
\begin{tikzpicture}[auto,x=1.4cm,y=.9cm, thick,minimum size=.8cm]

\node[draw,rectangle,rounded corners,minimum width = 12.2cm,minimum height=4.7cm] (spb) at (6.6,-1.6) {} ;

\node[MessageStartEvent] (jobe) at (1.5,0) {};
\node[lbl,below of=jobe,align = center] (jobe_name) {Job posted\\ $\left[\updatespec{InsJobCat}\right]$\\~};
\draw[sequence,->]  (jobe) -- (2.28,0);

\node[StartEvent] (se1) at (2.7,0) {};
\node[lbl,below of=se1,align = center] (se1_name) {};
\node[ExclusiveGateway,draw] (eg1) at (3.7,0) {};
\draw[sequence,->]  (se1) -- (eg1);

\node[EventBasedGateway,draw] (eg_err) at (4.8,0) {};
\node[draw,regular polygon,regular polygon sides=5,minimum width=1mm,scale=0.35] at (4.8,0){};
\draw[sequence,->]  (eg1) -- (eg_err);

\node[MessageIntermediateCatchingEvent] (appe) at (6,0) {};
\node[lbl ,below of=appe,align = center,yshift=1mm] (appe_name) {App.~received\\$\left[\updatespec{InsUser}\right]$};
\draw[sequence,->]  (eg_err) -- 
(appe);

\node[MessageIntermediateCatchingEvent,draw,minimum size=.8cm] (msg-exp) at (6,-2.2) {};
\node[lbl ,below of=msg-exp,align = center,yshift=1mm] (msg-exp_name) {Stop\\~};
\draw[sequence,rounded corners=5pt,->] (eg_err) |- 
(msg-exp);

\node[ErrorEndEvent,draw,minimum size=.8cm] (err) at (7.3,-2.2) {};
\node[lbl ,below of=err,align = center,yshift=1mm] (msg-exp_name) {Stopped\\~};
\draw[sequence,rounded corners=5pt,->] (msg-exp)--(err);


\node[task,align=center] (evalcv) at (7.3,0) {\tname{Evaluate}\\\tname{CV}};
\node[lbl ,below of=evalcv,align = center,yshift=1mm] {$\left[\updatespec{CheckQual}\right]$};
\draw[sequence,->]  (appe) -- (evalcv);

\node[ExclusiveGateway,draw] (eg2) at (8.6,0) {};
\draw[sequence,->]  (evalcv) -- (eg2);

\node[EndEvent] (ee1) at (10.5,0) {};
\node[lbl,below of=ee1,align = center] (ee1_name) {};
\draw[sequence,->]  (eg2) -- node[below,align=left] {$\cvar{qualif}=$\\\true} (ee1);

\node[task,align=center] (evalapp) at (8.6,-3.3) {\tname{Evaluate}\\\tname{Application}};
\node[lbl ,right of=evalapp,anchor=west,align = left] {$\left[\updatespec{EvalApp}\right]$};
\draw[sequence,->,rounded corners=5pt] (eg2) -- node[guard,align=left] {$\cvar{qualif}=$\\\false} (evalapp);
\draw[sequence,->,rounded corners=5pt] (evalapp) -| (eg1);

\node[ErrorIntermediateEvent,fill=white] (eb) at (10.95,-3.5) {};
\node[lbl,below right of=eb,align = center,xshift=-.05cm] (eb_name) {Stopped};

\node[task,align=center] (eligcand) at (13.3,-3.5) {\tname{Decide}\\\tname{Eligible}\\\tname{Candidates}};
\node[lbl ,right of=eligcand,anchor=west,align = left] {$\left[\updatespec{MarkE}\right]$};
\draw[sequence,->,rounded corners=5pt] (eb) -- (eligcand);

%
%

\node[task,align=center] (dwinner) at (13.3,-1.6) {\tname{Select}\\\tname{Winner}};
\node[lbl ,right of=dwinner,anchor=west,align = left] {$\left[\updatespec{SelWinner}\right]$};
\draw[sequence,->]  (eligcand) --  (dwinner);

%
%
\node[task,align=center] (awinner) at (12,0) {\tname{Assign}\\\tname{Winner}};
\draw[sequence,->,rounded corners=5pt] (10.93,0) -- (awinner);

\node[ExclusiveGateway,draw] (eg3) at (13.3,0) {};
\draw[sequence,->]  (awinner) --  (eg3);
\draw[sequence,->]  (dwinner) --  (eg3);

\node[task,align=center] (offer) at (14.5,0) {\tname{Make}\\\tname{Offer}};
\draw[sequence,->]  (eg3) --  (offer);

\node[EndEvent] (ee2) at (15.9,0) {};
\node[lbl,below of=ee2,align = center] (ee2_name) {};
\draw[sequence,->]  (offer) --  (ee2);

\end{tikzpicture}
}
\caption{The job hiring process. Elements in squared brackets indicate the update specifications attached to the corresponding tasks/events, and formalized as shown in Examples~\ref{ex:updates-1} and~\ref{ex:updates-2}.}
\label{fig:bpmn-hiring}
\end{figure}
\subsection{Execution Semantics}
\label{sec:exec}
We intuitively describe the execution semantics of a \dab $\M = \tup{\D,\P}$, using the update/task logic and progression rules of blocks as a basis. Upon execution, each state of $\M$ is characterized by an \emph{$\M$-snapshot}, in turn constituted by a data snapshot of $\D$ (cf.~Section~\ref{sec:data-schema}), and a \emph{control map} whose keys are the (active) case identifiers, and whose values are assignments from each block in $\P$ to one of its lifecycle states.

Initially, the data snapshot fixes the immutable content of the catalog $\dcat{\D}$, while the repository instance, the case map, and the control map, are all empty. At each moment in time, the $\M$-snapshot is then evolved by nondeterministically performing one of the two steps:
\begin{compactdesc}
\item[(new case creation)] A new case is created, obtaining a corresponding fresh identifier $id$. The new $\M$-snapshot is then obtained by maintaining the catalog and repository unaltered, while updating:
\begin{inparaenum}[\it (i)] 
\item the case map, creating a new entry with key $id$ and setting all the case variables to $\nullv$ in such a newly created entry, with the exception of $\self$, which gets the fresh identifier $id$;
\item the control map, creating a new entry with key $id$ and setting all the block lifecycles to $\sidle$ in such a newly created entry. 
\end{inparaenum}
\item[(case progression)] The identifier $id$ of a case present in the control/case map is nondeterministically picked, nondeterministically evolving it of one step through the process, depending on the current $\M$-snapshot. The new $\M$ -snapshot is then obtained by manipulating, accordingly, the $id$ entries in the case and control maps, as well as possibly updating the repository if the selected step involves the application of an effect rule (cf.~Section~\ref{sec:update-logic}). 
\end{compactdesc}
 
\newcommand{\tsys}[3]{\Upsilon^{#3}_{#1,#2}}
\newcommand{\bralgo}{\mathsf{BackReach}}

\section{Parameterized Verification of Safety Properties}
\label{sec:verification}
We now focus on parameterized verification of \dab{s} using SMT-based techniques grounded in the theory of arrays. 

\subsection{Array-Based Artifact Systems and Safety Checking}
We recall the key notions behind array-based systems, and the artifact variants recently studied in \cite{CGGMR18,CGGMR19} to bridge the gap between SMT-based model checking of array-based systems  \cite{ijcar08,lmcs} and  verification of artifact-centric processes \cite{Vian09,DeLV16}. 
In general terms, an array-based system is described using a multi-sorted
theory that contains two types of sorts, one accounting for the indexes of
arrays, and the other for the elements stored therein. Since the content of an array changes over time, it is referred to using a \emph{function}
variable, whose interpretation in a state is that of a total function mapping indexes to elements (so that applying the function to an index denotes the classical \emph{read} operation for arrays). The definition of an array-based system with array state variable $a$ always requires
\begin{inparaenum}[\it (i)]
\item a formula $I(a)$ describing the \emph{initial configuration} of the array $a$;
\item  a formula $\tau(a,a')$ describing a \emph{transition} that transforms the content of the
array from $a$ to $a'$. 
\end{inparaenum}
In such a setting, verifying whether the system can
\emph{reach} unsafe configurations described by a formula $K(a)$ amounts to check
whether the formula
$I(a_0)\wedge \tau(a_0, a_1) \wedge \cdots \wedge \tau(a_{n-1}, a_n)\wedge
K(a_n)$ is satisfiable for some $n$.
Notably, several mature model checkers exist to ascertain safety of these type of systems, such as \textsc{mcmt} \cite{mcmt} 
and \textsc{cubicle}~\cite{cubicle_cav}. 
In \cite{CGGMR18,CGGMR19}, we have extended array-based systems towards an array-based version of the artifact-centric model, considering two main settings:
\begin{compactdesc}
  \item[(simple artifact systems - SAS)] artifact systems operating over a read-only relational database that resembles our \dab catalog, and over a single tuple (or boundedly many tuples) of updatable elements;  
  \item[(relational artifact systems - RAS)] systems that extend SAS with a relational storage for unboundedly many updatable elements. 
\end{compactdesc}
Notably, safety properties are checked over such systems irrespectively of the content of the read-only database, following the tradition of \cite{Vian09,DeLV16}. Several soundness, completeness, and decidability results have been obtained by suitably controlling the expressiveness of these systems. In addition, from Version~2.8 \textsc{mcmt} has been extended to handle safety checking of SAS and RAS.

\subsection{Verification Problems for \dab{s}}
First of all, we need a language to express undesired properties over a \dab $\M = \tup{\D,\P}$. To do so, we resort to the same \emph{guard} language introduced in Definition~\ref{def:guard}, extended with two features. First, we support  querying also the control map of each case, so as to express control-flow properties. This is done by simply extending $\D$ with additional, special case control variables referring to the lifecycle state of the blocks in $\P$ (where each block $B$ gets variable $\cvar{Blifecycle}$). Each such variable is assigned, in a given snapshot, to the state of the corresponding block lifecycle (i.e., $\sidle$, $\senabled$, and the like). We call thes additional case variables $\mathit{F}_\P$. Second, we allow the modeler to inspect the status of multiple cases at once, so as to check how different cases may implicitly influence each other via the global repository. This is done by ``indexing" guards, with the assumption that different indexes denote different cases. 

\begin{definition}
Let $I$ be a set of $n$ distinct indexes. A \emph{property} over $n$ cases of $\M = \tup{\D,\P}$ is a formula of the form $\bigwedge_{i \in I} G_i[i]$, where each $G_i$ is a guard over $\D$ and the case control variables $\mathit{F}_\P$ satisfying the following condition: every non-case variable $x \in \getdatavars{G_i}$ appearing in $G_i$ must also appear in an atom whose relation schema belongs to the repository $\drepo{\D}$.
\end{definition}

\begin{example}
By calling $HP$ the root, process block of Figure~\ref{fig:bpmn-hiring}, we can test whether some case of the process can terminate through the property $(\cvar{HPlifecycle} = \constant{completed})[i]$. If the property is unreachable, then no case can be progressed from the start to the end of the process. Since \dab processes are block structured, this is enough to ascertain whether the process is \emph{unsound}.
\end{example}

%

We study (un)safety of these properties by considering general, unrestricted \dab{d}, and also two interesting fragments:
\begin{compactdesc}
\item[(case-bounded)] \dab{s} that only introduce a bounded number of cases during their execution, where the bound is known a-priori - a typical setting being the one where a single case is studied, in the style of soundness verification for workflow nets;  
\item[(repo-bounded)] \dab{s} that introduce only boundedly many tuples in the repository during  their execution, where the bound is known a-priori.     
\end{compactdesc}
In this light, verification of safety properties over \dab{s} present multiple dimensions of parameterization, depending on the type of \dab under analysis. The two extremes are: 
\begin{inparaenum}[\it (i)]
\item   
case- and repo-bounded \dab{s}, for which the only parameter is the instantiation of the catalog;
\item unrestricted \dab{s}, where verification is carried out parametrically w.r.t.~the instantiation of the catalog, the number of tuples in the repository, and the overall number of cases. 
\end{inparaenum}

\subsection{Translating \dab{s} into Array-Based Artifact Systems}
\label{sec:translation}
To attack parameterized verification problems over \dab{s}, we translate them into corresponding verification problems over SAS and RAS. We only provide here the main intuitions behind the translation, which is fully addressed in the Appendix. Let $\M = \tup{\D,\P}$ be a \dab. $\dcat{\D}$ is maintained unaltered, as it is addressed in SAS and RAS in its full generality. The translation of $\dcvars{\D}$ and $\drepo{\D}$ depends instead on whether $\M$ is studied unrestrictedly, or under case- and/or repo-boundedness. Consider $\drepo{\D}$. If $\M$ is repo-bounded with a bound of $1$, then every relation schema in $\drepo{\D}$ has just one tuple, and consequently can be represented using a set of (global) variables, one per relation attribute, in the style of a SAS. A $k$-bounded setting is handled similarly, just replicating the variables for $k$ times. If instead no boundedness assumption on the report is placed, for each relation schema $R \in \drepo{\D}$, and each attribute $a \in \relattrs{R}$, a dedicated array is introduced. The index of the array represents the (implicit) identifier of $R$, in line with our repository model. To reconstruct a specific tuple from $R$, one just needs to retrieve the objects present in the arrays corresponding to the different attributes of $R$, always using the same index $i$. The resulting model corresponds to that of a RAS and its notion of \emph{artifact relation} \cite{CGGMR18,CGGMR19}. 

A similar strategy is adopted for the case variables: if no bound on the number of cases is given, then each case variable is translated into a corresponding array, whose elements maintain the value that one case is assigning to that variable. Accessing all such arrays with the same index produces back the entire case variable assignments for the corresponding case. Finally, $\self$ is handled by introducing an array with the property that its elements are in bijection with the indexes (i.e., no element repeats twice in the array). Exactly the same approach is replicated to store the control information about blocks on a per-case basis. 

All in all, depending on the boundedness assumptions on cases and/or repository, the translation produces a SAS or a RAS with different artifact relations. Each transition formula realizes one of the progression rules that collectively realize the execution semantics of the input \dab (cf.~Section~\ref{sec:exec}).

 In \cite{CGGMR18,CGGMR19}, we focus on parameterized (un)safety of RAS, verifying whether there exists an instance of the read-only database such that the artifact system can reach an unsafe configuration.  Since the cells of the arrays may point to identifiers in the catalog, in turn related to other catalog relations via foreign keys, the standard backward reachability procedure needs to be suitably revised \cite{CGGMR19}. In fact, when computing preimage formulae over RAS, existentially quantified ``data" variables may be introduced, breaking the format of state formulae. To restore the key property that the preimage of a state is again represented symbolically as a state formula, such additional quantified variables must be eliminated. Suitable quantifier elimination techniques have been studied in \cite{CGGMR19,cade19} and implemented in the latest version~2.8 of \mcmt, which can now natively handle the verification of RAS. In addition, while the unsafety verification is in general undecidable for RAS, several subclasses with decidable unsafety have been singled out. One of such classes corresponds to RAS operating over arrays whose maximum size is bounded a-priori, i.e. SAS. All in all, the RAS framework provides a natural foundational and practical basis to formally analyze \dab{s}, which we tackle next.

From now on, we use $\bralgo$ to refer to the backward reachability procedure that:
\begin{compactitem} 
\item takes as input a \dab, a property to be verified, and a series of additional information related to the boundedness assumptions and the adopted semantics for insertion (set vs multiset);
\item translates the input \dab into a corresponding SAS/RAS (according to the provided additional information), and the input property into a corresponding property over the target SAS/RAS;
\item invokes the SAS/RAS backward reachability procedure described in \cite{CGGMR18,CGGMR19} and implemented in \textsc{mcmt};
\item returns \emph{yes} if and only if the property is reachable.
\end{compactitem}

All the proofs of the following theorems are obtained by exploiting the translation, and by showing that the SAS/RAS produced from the translation enjoys the property stated in the theorem.

\subsection{Soundness and Completeness Results}
\label{sec:soundness-completeness}
We start by considering case-bounded systems.
\begin{theorem}
\label{thm:case-bounded}
  $\bralgo$ is sound and complete for case-bounded \dab{s} that use the multiset or set insertion semantics.  
\end{theorem}
While for case-bounded \dab{s} soundness and completeness are guaranteed without additional restrictions, this is not the case in the unrestricted setting. The problem is, in fact, the usage of $\self$, which implicitly allows to create references across read-write relations (something that is not allowed in a \dab repository, nor in the corresponding model of RAS). We recover soundness and completeness by disallowing the explicit usage of $\self$.

\begin{definition}
A \dab $\M$ is \emph{case-identifier-agnostic} if none of the update specifications in $\M$ mentions $\self$.
\end{definition}

\begin{theorem}
\label{thm:unrestricted}
  $\bralgo$ is sound and complete for case-identifier-agnostic \dab{s} that use the multiset or set insertion semantics.    
\end{theorem}
We stress here that soundness and completeness indicate that whenever $\bralgo$ terminates, it produces a correct answer. Termination is not guaranteed in the general case (but may very well be obtained on the analyzed \dab), and consequently $\bralgo$ is a semi-decision procedure.

\subsection{Termination Results}
\label{sec:termination}
We now discuss how the previous results can be strengthen to ensure termination (thus witnessing decidability of parameterized verification). The first, unavoidable limitation that we have to impose is on the constraints used in the catalog: its foreign keys cannot form cycles. This is in line with \cite{verifas,CGGMR19}. To define acyclicity, we associate a catalog $\cat$ to a characteristic graph $G(\cat)$ that captures the dependencies between relation schema components induced by primary and foreign keys. Specifically,  $G(\cat)$ is a directed graph such that:
\begin{compactitem}[$\bullet$]
\item for every $R \in \cat$ and every attribute $a \in \relattrs{R}$, the pair $\tup{R,a}$ is a node of $G(\cat)$ (and nothing else is a node); 
\item $\tup{R_1,a_1} \rightarrow \tup{R_2,a_2}$ is and only if one of the two cases apply:
\begin{inparaenum}[\it (i)]
\item  $R_1=R_2$, $a_2 \neq a_1$, and $a_1 = \relid{R}$;
\item  $a_2 = \relid{R_2}$ and $a_1$ is a foreign key referring $R_2$. 
\end{inparaenum}
\end{compactitem}

\begin{definition}
A \dab is acyclic if the characteristic graph of its catalog is so.
\end{definition}

\begin{theorem}
\label{thm:dec-case-repo-bounded}
   $\bralgo$ terminates when verifying properties over case- and repo-bounded, acyclic \dab{s} using the multiset or set insertion semantics.  
\end{theorem}
This strong result is obtained due to the fact that case- and repo-bounded \dab{s} get translated into SAS, where the read-write storage is constituted by a fixed set of variables. If instead we consider more sophisticated \dab{s} that get translated into RAS with their sophisticated read-write relational storage, then termination requires to carefully control the interplay between the different components of the \dab. While the required conditions are quite difficult to grasp at the syntactic level, they can be intuitively understood as follows: to ensure termination, whenever the progression of the \dab depends on the repository, it does so only via a single entry in one of its relations. This indicates that direct or indirect comparisons and joins of distinct tuples within the same or different repository relations cannot be used in an update. Towards avoiding indirect joins, queries cannot that mix case variables and repository relations, nor update case variables with the content of other case variables. The following definition is instrumental to enforce this principle.

\begin{definition}
  \label{def:separated-guard}
  A \emph{guard} $G \triangleq q(\vec{x}) \leftarrow \bigvee_{i=1}^n Q_i$ over
  a data component $\D$ is \emph{separated} if, for every $i$, $j$ we have that
  $\getdatavars{Q_i}\cap \getdatavars{Q_j}=\emptyset$ and each $Q_i$ is of the
  form $\chi \land R(\vec{y}) \land \xi$ (here, $\chi$, $R(\vec{y})$ or $\xi$
  are optional), where:
  \begin{inparaenum}[\itshape (i)]
  \item $\chi$ is a conjunctive query with filters over $\dcat{\D}$ only (that
    can employ case variables);
  \item $R \in \drepo{\D}$ is a repository relation schema;
  \item $\vec{y}$ is a tuple of variables and/or constant objects in
    $\datadom$, such that $\vec{y} \cap \dcvars{\D} = \emptyset$, and
    $\getdatavars{\chi}\cap \vec{y}=\emptyset$;
  \item $\xi$ is a conjunctive query with filters over $\dcat{\D}$ only, that
    possibly mentions variables in $\vec{y}$ but does \emph{not} include any
    case variable (i.e., $\getcasevars{\xi} = \emptyset$), and
    such that $\getdatavars{\chi}\cap\getdatavars{\xi}=\emptyset$.
  \end{inparaenum}
  A \emph{property} is \emph{separated} if all its inner guards are separated.
\end{definition}

Intuitively, a separated guard consists of two isolated parts: one part $\chi$
inspecting the content of case variables and their relationship with the
catalog, and another part $R(\vec{y}) \land \xi$ retrieving a single tuple
$\vec{y}$ in some repository relation $R$, possibly filtering it through
inspection of the catalog via $\xi$.

\begin{example}
  Consider the refinement
  $\getpre{\updatespec{EvalApp}} \triangleq
  \relname{GetScore}(s:\sort{NumScore}) \leftarrow \xi \land \chi$ of the guard
  $\getpre{\updatespec{EvalApp}}$ from Example~\ref{ex:updates-1}, where
  $\chi:=\relname{User}(\cvar{uid},name,age)$ checks if the
  variables $\tup{\cvar{uid},name,age}$ form a tuple in
  $\relname{User}$, and $\xi:=1 \leq s \land s \leq 100$. This guard is
  separated since $\chi$ and $\xi$ match the requirements of the previous
  definition.
 \end{example}

\begin{theorem}
\label{thm:dec-case-bounded}
Let $\M$ be a case-bounded, acyclic \dab that uses the multiset insertion semantics, and is so that for each update specification $\updatespec{u}$ of $\M$, the following holds:  \begin{compactitem}[$\bullet$]
  \item
  If $\getpost{\updatespec{u}}$ is an \emph{insert\&set} rule (with an explicit $\INSERT$ part), then  $\getpre{\updatespec{u}}$ is \emph{repo-free}; 
  \item If $\getpost{\updatespec{u}}$ is a \emph{set} rule (not containing an $\INSERT$ part), then either
  \begin{inparaenum}[\it (i)]
    \item $\getpre{\updatespec{u}}$ is \emph{repo-free}, or
    \item $\getpre{\updatespec{u}}$ is \emph{separated} and \emph{all} case variables $\dcvars{\D}\setminus \set{\self}$ appear in the $\SET$ part of $\getpost{\updatespec{u}}$;   
  \end{inparaenum}
  \item If $\getpost{\updatespec{u}}$ is a \emph{delete\&set} rule, then $\getpre{\updatespec{u}}$ is \emph{separated} and \emph{all} case variables $\dcvars{\D}\setminus \set{\self}$ appear in the $\SET$ part of $\getpost{\updatespec{u}}$;  
  \item If $\getpost{\updatespec{u}}$ is a \emph{conditional update} rule, then  $\getpre{\updatespec{u}}$ is \emph{repo-free} and \emph{boolean}, so that $\getpost{\updatespec{u}}$ only makes use of the new variables introduced in its $\UPDATE$ part (as well as constant objects in $\datadom$).
  \end{compactitem}
Then, $\bralgo$ terminates when verifying separated properties over $\M$.
\end{theorem}
%
\begin{theorem}
\label{thm:dec-case-unbounded}
   Let $\M$ be a case-identifier-agnostic acyclic \dab that uses the multiset insertion semantics, and is so that for each update specification $\updatespec{u}$ in $\M$, $\updatespec{u}$ satisfies the same conditions as of Theorem~\ref{thm:dec-case-bounded}.
   Then, $\bralgo$ terminates when verifying separated properties over $\M$.
\end{theorem}
It is important to notice that the conditions in Theorems~\ref{thm:dec-case-bounded} and \ref{thm:dec-case-unbounded} represent a concrete, BPMN-like counterpart of the decidability results in \cite{verifas} and \cite{CGGMR19}.

\begin{example}
  Our hiring job \dab makes use of $\self$ towards relating applications to the identifier of the case to which they were submitted. Hence, if we want to retain soundness and completeness of $\bralgo$, we have to restrict the analysis to the case-bounded setting. By considering the data and process schema of the \dab, we can directly show that it obeys to all conditions in Theorem~\ref{thm:case-bounded}, in turn guaranteeing termination of $\bralgo$.
  \end{example}

\section{First Experiments with MCMT}
\label{sec:mcmt}
We have manually encoded the job hiring \dab described in the paper as an \mcmt specification, using the same translation rules recalled in Section~\ref{sec:translation} and fully spelled out in the Appendix towards proving the main theorems in Sections~\ref{sec:soundness-completeness} and \ref{sec:termination}. \textsc{mcmt} 
checks unsafety of a property through a symbolic, backward reachability procedure. 
The algorithm computes iterated preimages of the given property and applies to
them quantifier elimination, until a fixpoint is reached or until a set
intersecting the initial state (i.e., also characterized using a formula) is found. To do this efficiently, \mcmt is equipped with dedicated quantifier elimination techniques, and discharge safety and fixpoint tests encountered during the backward search to state-of-the-art SMT solvers.

We have checked the encoding of the process  against five safe and five unsafe properties. The first property ascertains whether a job hiring case can actually reach the end point of the process, in turn witnessing soundness. The BPM\_SAFE1 property checks whether it is possible to have a situation where the \tname{Select Winner} task is $\senabled$ and the case variable $\cvar{result}$ indicates that the winner is not eligible.

 Table~\ref{tab:exp-results} shows the so-obtained, very encouraging results. Such initial results nicely complement those carried out on RAS in \cite{CGGMR18}, indicating that the approach is promising not just foundationally, but also in terms of tool support. These experiments, together with the ones reported in this paper, are available as part of the last distribution~2.8 of
\textsc{mcmt}.\footnote{\url{http://users.mat.unimi.it/users/ghilardi/mcmt/},
 subdirectory \texttt{/examples/dbdriven} of the distribution. The user manual
 contains a new section (pages 36--39) on how to encode \ras{s} in MCMT
 specifications.} 
 Experiments were performed on a machine with Ubuntu~16.04, 2.6\,GHz Intel
Core~i7 and 16\,GB RAM.

\begin{table}[t!]
  \begin{tabularx}{\textwidth}{lXXlXX}
\footnotesize
    \textbf{Exp.}  & \textbf{Res.} & \textbf{Time (s)} & 
    \textbf{Exp.}  & \textbf{Res.} & \textbf{Time (s)} 
    \\\midrule
    BPM\_end\_process 
          & \unsafe   &  0.43
    \\
    BPM\_SAFE1  & \safe & 0.20 
    &
    BPM\_UNSAFE1 & \unsafe   & 0.18
    \\
    BPM\_SAFE2  & \safe & 5.85 
    &
    BPM\_UNSAFE2 & \unsafe & 1.17 
    \\
    BPM\_SAFE3 & \safe & 3.56 
    &
    BPM\_UNSAFE3 & \unsafe   & 4.45 
    \\
    BPM\_SAFE4  & \safe   & 0.03 
    &
    BPM\_UNSAFE4 & \unsafe & 1.43
    \\
    BPM\_SAFE5 &  \safe & 0.27 
    &
    BPM\_UNSAFE5 & \unsafe   & 1.14
    \\    
    \bottomrule
  \end{tabularx}
  \vspace{3pt}
\caption{Time spent by \mcmt to check different properties over the job hiring \dab. Names of experiments coincide with those of the \mcmt files from the Ancillary files of arXiv:1905.12991.}
\label{tab:exp-results}
\end{table}

\section{Conclusion}\label{sec:conclusion}
In this paper, we have introduce a data-aware version of BPMN, called \dab, that achieves an interesting trade-off between expressiveness, and the possibility of applying sophisticated parameterized verification techniques to ascertain safety of the produced models. In particular, we have identified classes of \dab{s} for which backward reachability techniques coming from the SMT tradition are correct, further strengthening them to also guarantee termination of backward reachability.
From the foundational point of view, we are interested in equipping \dab{s} with datatypes and corresponding rigid predicates, including arithmetic operators, as done in \cite{DeLV16} for artifact-centric systems. This is promising especially considering that there are plenty of state-of-the-art SMT techniques to handle arithmetics. At the same time, we want to attack the main limitation of our approach, namely that guards and conditions are actually existential formulae, and the only (restricted) form of universal quantification available in the update language is that of conditional updates. Universal guards in transition formulae could be very useful in specifications: for example, they would allow us to specify a branch in a job hiring process that is followed only if no applicant satisfies a certain condition. This is reminiscent to what happens in the  verification of distributed systems, where universal guards frequently occur in specifications. The question has been debated since longtime in the literature and the most effective solution adopted to cope with this problem so far is the introduction of suitable ``monotonic abstractions" (see \cite{AlGS14} for a survey). Notably, this solution is currently implemented in \mcmt.
 Monotonic abstractions could introduce spurious unsafe traces, and in fact \mcmt warns the user about this (in practice, not so frequent) possibility. 
 
 An orthogonal, challenging question is how, and to what extent, some of the most recent techniques developed for temporal model checking of artifact-centric systems \cite{DeLV16} can be incorporated in our approach, allowing us to prove more sophisticated properties beyond safety.



From the experimental point of view, while a systematic evaluation is out of scope of this paper, the initial experiments carried out in this paper and \cite{CGGMR18} indicate that the approach is promising. We intend to fully automate the translation from \dab{s} to array-based systems, and to set up a benchmark to evaluate the performance of verifiers for data-aware processes, starting from the examples collected in \cite{verifas} (which are inspired from BPMN processes, and consequently should be straightforwardly encoded ad \dab{s}). 

\bibliographystyle{abbrv}
\bibliography{main-bib}

\newpage
\appendix
\section{\dab Blocks}
\label{app:blocks}
\newcommand{\legendw}{8cm}
\newcommand{\bdist}{5mm}


\subsection{Basic Blocks}
\label{sec:basic}

\resizebox{\textwidth}{!}{
\begin{tikzpicture}[x=1.8cm,y=.9cm, thick]

\matrix[  nodes={node distance=\bldist},
            rectangle,
            nodes in empty cells,
            row sep=.5mm,column sep=1mm,
            very thick,
            column 1/.style={anchor=west},
            column 2/.style={anchor=west},
            column 3/.style={anchor=west},
            >=latex,->,
            ampersand replacement=\&
          ] (declarematrix) {
\node{\textbf{Block}};
\&
\&
\node{\textbf{Attributes}};
\\
\node{empty};
\&
  \node[smalltask,white] (task) {};
  \draw[outflow,thin] (task.west) -- (task.east);
\&
\node{};
\\
\node{task};
\&
  \node[smalltask] (task) {\textsf{A}};
  \draw[outflow,thin] ($(task.west)-(\outfdist,0)$) -- (task);
  \draw[outflow,thin] (task) -- ($(task.east)+(\outfdist,0)$);
\&
\node{
\begin{tabular}{@{}p{\legendw}@{}}
(1) Atomic/non-atomic\\
(2) update specification.
\end{tabular}
};
\\
\node{catch event};
\&
  \node[IntermediateEvent,minimum size=.6cm] (task) {$e$};
  \draw[outflow,thin] ($(task.west)-(\outfdist,0)$) -- (task);
  \draw[outflow,thin] (task) -- ($(task.east)+(\outfdist,0)$);
\&
\node{
\begin{tabular}{@{}p{\legendw}@{}}
(1) Type of event $e$ (msg, timer, none)\\
(2) update specification.
\end{tabular}
};
\\
\node{process block};
\&
\node[StartEvent,minimum size=.6cm] (start) {$e_s$};
\node[smallblock,right=\bdist of start] (task) {\textsf{B}};
\draw[sequence,->] (start) -- (task);
\node[EndEvent,minimum size=.6cm,right= 5mm of task] (end) {$e_t$};
\draw[sequence,->]  (task) -- (end);
\&
\node{
\begin{tabular}{@{}p{\legendw}@{}}
(1) Type of start event $e_s$ (msg, timer, none)\\
(2) Update specification of $e_s$\\
(3) Type of end event $e_t$ (msg, none)\\
(4) Update specification of $e_t$\\
(5) Arbitrary nested block $\textsf{B}$
\end{tabular}
};
\\
\node{subprocess};
\&
\node[smalltask] (task) at (0,0) {};
\node[below=0mm of task.north,anchor=north] {\textsf{A}};
\node[
  rectangle,
  anchor=south,
  above=0mm of task.south,
  draw,
  minimum size=.5mm,
] {\tiny +};
  \draw[outflow,thin] ($(task.west)-(\outfdist,0)$) -- (task);
  \draw[outflow,thin] (task) -- ($(task.east)+(\outfdist,0)$);
\&
\node{
\begin{tabular}{@{}p{\legendw}@{}}
(1) Inner process block
\end{tabular}
};
\\
};
\end{tikzpicture}
}

\clearpage
\subsection{Flow Blocks}
\label{sec:flow-blocks}
For simplicity, we consider only two nested blocks, but multiple nested blocks can be seamlessly handled.

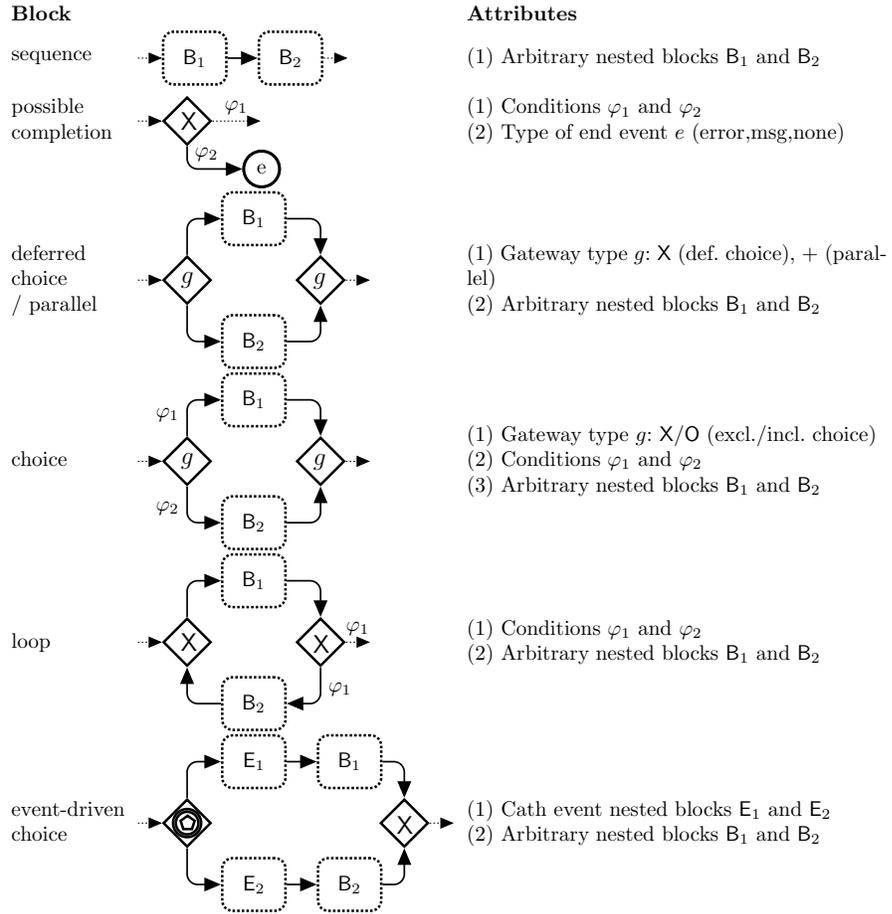
\renewcommand{\legendw}{7cm}
\resizebox{\textwidth}{!}{
\begin{tikzpicture}[x=1.8cm,y=.9cm, thick]

\matrix[  nodes={node distance=\bldist},
            rectangle,
            nodes in empty cells,
            row sep=.5mm,column sep=1mm,
            very thick,
            column 1/.style={anchor=west},
            column 2/.style={anchor=west},
            column 3/.style={anchor=west},
            >=latex,->,
            ampersand replacement=\&
          ] (declarematrix) {
\node{\textbf{Block}};
\&
\&
\node{\textbf{Attributes}};
\\
\node{sequence};
\&
\node[smallblock] (task1) {\textsf{B}$_1$};
\node[smallblock,right=\bdist of task1] (task2) {\textsf{B}$_2$};
\draw[outflow,thin] ($(task1.west)-(\outfdist,0)$) -- (task1);
\draw[sequence,->] (task1) -- (task2);
\draw[outflow,thin] (task2) -- ($(task2.east)+(\outfdist,0)$);
\&
\node{
\begin{tabular}{@{}p{\legendw}@{}}
(1) Arbitrary nested blocks $\textsf{B}_1$ and $\textsf{B}_2$
\end{tabular}
};
\\
\node{\begin{tabular}{@{}l@{}}possible\\completion\end{tabular}};
\&
\node[ExclusiveGateway,minimum size=8mm] (eg) {};
\node {~\,\large\textsf{X}};
\draw[outflow,thin] ($(eg.west)-(\outfdist,0)$) -- (eg);
\draw[outflow,thin,->] (eg) -- node[above] {$\varphi_1$} ($(eg.east)+(2*\outfdist,0)$);
\node[EndEvent,draw,minimum size=.6cm,right=\bdist of eg,yshift=-8mm] (err) {e};
\draw[sequence,->,rounded corners=5pt] (eg) |- node[above right] {$\varphi_2$} (err);
\&
\node{
\begin{tabular}{@{}p{\legendw}@{}}
(1) Conditions $\varphi_1$ and $\varphi_2$\\
(2) Type of end event $e$ (error,msg,none)
\end{tabular}
};
\\
\node{\begin{tabular}{@{}l@{}}deferred\\choice\\ / parallel\end{tabular}};
\&
\node[ExclusiveGateway,minimum size=8mm] (eg) {};
\node {~\,\large{$g$}};
\draw[outflow,thin] ($(eg.west)-(\outfdist,0)$) -- (eg);
\node[smallblock,above right=\bdist of eg] (task1) {\textsf{B}$_1$};
\node[smallblock,below right=\bdist of eg] (task2) {\textsf{B}$_2$};
\node[ExclusiveGateway,minimum size=8mm,below right=\bdist of task1] (eg2) {};
\node[below right=\bdist of task1,xshift=-.5mm,yshift=.5mm] {\large{$g$}};
\draw[outflow,thin] (eg2) -- ($(eg2.east)+(\outfdist,0)$);
\draw[sequence,->,rounded corners=5pt] (eg) |- (task1);
\draw[sequence,->,rounded corners=5pt] (eg) |- (task2);
\draw[sequence,->,rounded corners=5pt] (task1) -| (eg2);
\draw[sequence,->,rounded corners=5pt] (task2) -| (eg2);
\&
\node{
\begin{tabular}{@{}p{\legendw}@{}}
(1) Gateway type $g$: $\textsf{X}$ (def.~choice), $\textsf{+}$ (parallel)\\
(2) Arbitrary nested blocks $\textsf{B}_1$ and $\textsf{B}_2$
\end{tabular}
};
\\
\node{\begin{tabular}{@{}l@{}}choice\end{tabular}};
\&
\node[ExclusiveGateway,minimum size=8mm] (eg) {};
\node {~\,\large{$g$}};
\draw[outflow,thin] ($(eg.west)-(\outfdist,0)$) -- (eg);
\node[smallblock,above right=\bdist of eg] (task1) {\textsf{B}$_1$};
\node[smallblock,below right=\bdist of eg] (task2) {\textsf{B}$_2$};
\node[ExclusiveGateway,minimum size=8mm,below right=\bdist of task1] (eg2) {};
\node[below right=\bdist of task1,xshift=-.5mm,yshift=.5mm] {\large{$g$}};
\draw[outflow,thin] (eg2) -- ($(eg2.east)+(\outfdist,0)$);
\draw[sequence,->,rounded corners=5pt] (eg) |- node[below left] {$\varphi_1$} (task1);
\draw[sequence,->,rounded corners=5pt] (eg) |- node[above left] {$\varphi_2$} (task2);
\draw[sequence,->,rounded corners=5pt] (task1) -| (eg2);
\draw[sequence,->,rounded corners=5pt] (task2) -| (eg2);
\&
\node{
\begin{tabular}{@{}p{\legendw}@{}}
(1) Gateway type $g$: $\textsf{X}$/$\textsf{O}$ (excl./incl.~choice)\\
(2) Conditions $\varphi_1$ and $\varphi_2$\\
(3) Arbitrary nested blocks $\textsf{B}_1$ and $\textsf{B}_2$
\end{tabular}
};
\\
\node{loop};
\&
\node[ExclusiveGateway,minimum size=8mm] (eg) {};
\node {~\,\large{\textsf{X}}};
\draw[outflow,thin] ($(eg.west)-(\outfdist,0)$) -- (eg);
\node[smallblock,above right=\bdist of eg] (task1) {\textsf{B}$_1$};
\node[smallblock,below right=\bdist of eg] (task2) {\textsf{B}$_2$};
\node[ExclusiveGateway,minimum size=8mm,below right=\bdist of task1] (eg2) {};
\node[below right=\bdist of task1,xshift=-.5mm,yshift=.5mm] {\large{\textsf{X}}};
\draw[outflow,thin] (eg2) -- node[above] {$\varphi_1$} ($(eg2.east)+(\outfdist,0)$);
\draw[sequence,->,rounded corners=5pt] (eg) |- (task1);
\draw[sequence,->,rounded corners=5pt] (task2) -| (eg);
\draw[sequence,->,rounded corners=5pt] (task1) -| (eg2);
\draw[sequence,->,rounded corners=5pt] (eg2) |- node[above right] {$\varphi_1$} (task2);
\&
\node{
\begin{tabular}{@{}p{\legendw}@{}}
(1) Conditions $\varphi_1$ and $\varphi_2$\\
(2) Arbitrary nested blocks $\textsf{B}_1$ and $\textsf{B}_2$
\end{tabular}
};
\\
\node{\begin{tabular}{@{}l@{}}event-driven\\choice\end{tabular}};
\&
\node[EventBasedGateway,draw,minimum size=8mm] (eg) {};
\node[draw,regular polygon,regular polygon sides=5,minimum width=2mm,scale=0.6,xshift=5.2mm]{};
\draw[outflow,thin] ($(eg.west)-(\outfdist,0)$) -- (eg);
\node[smallblock,above right=\bdist of eg] (e1) {\textsf{E}$_1$};
\node[smallblock,below right=\bdist of eg] (e2) {\textsf{E}$_2$};
\node[smallblock,right=\bdist of e1] (task1) {\textsf{B}$_1$};
\node[smallblock,right=\bdist of e2] (task2) {\textsf{B}$_2$};
\node[ExclusiveGateway,minimum size=8mm,below right=\bdist of task1,xshift=-2mm] (eg2) {};
\node[below right=\bdist of task1,xshift=-2.5mm,yshift=.5mm] {\large{\textsf{X}}};
\draw[outflow,thin] (eg2) -- ($(eg2.east)+(\outfdist,0)$);
\draw[sequence,->,rounded corners=5pt] (eg) |- (e1);
\draw[sequence,->,rounded corners=5pt] (eg) |- (e2);
\draw[sequence,->,rounded corners=5pt] (e1) -- (task1);
\draw[sequence,->,rounded corners=5pt] (e2) -- (task2);
\draw[sequence,->,rounded corners=5pt] (task1) -| (eg2);
\draw[sequence,->,rounded corners=5pt] (task2) -| (eg2);
\&
\node{
\begin{tabular}{@{}p{\legendw}@{}}
(1) Cath event nested blocks $\textsf{E}_1$ and $\textsf{E}_2$\\
(2) Arbitrary nested blocks $\textsf{B}_1$ and $\textsf{B}_2$
\end{tabular}
};
\\
};
\end{tikzpicture}
}

\clearpage
\subsection{Exception Handling Blocks}
\label{sec:exception-blocks}

For simplicity, we show a single boundary event, but multiple boundary events and their corresponding handlers can be seamlessly handled.

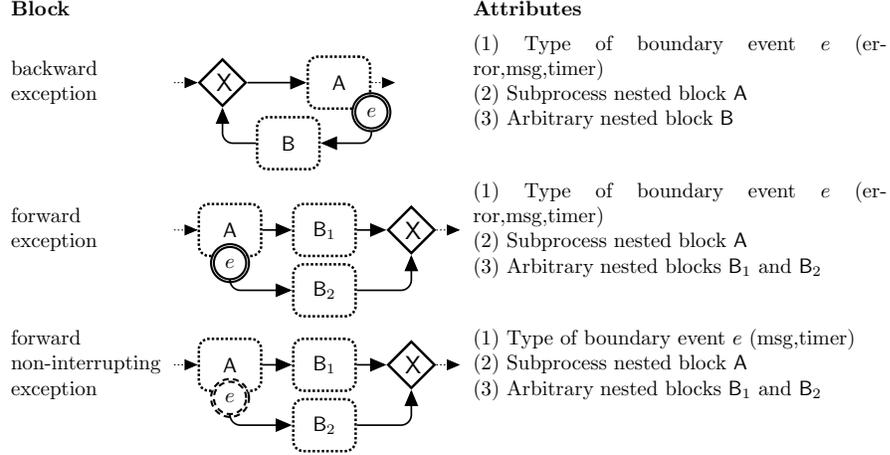
\resizebox{\textwidth}{!}{
\begin{tikzpicture}[x=1.8cm,y=.9cm, thick]

\matrix[  nodes={node distance=\bldist},
            rectangle,
            nodes in empty cells,
            row sep=.5mm,column sep=1mm,
            very thick,
            column 1/.style={anchor=west},
            column 2/.style={anchor=west},
            column 3/.style={anchor=west},
            >=latex,->,
            ampersand replacement=\&
          ] (declarematrix) {
\node{\textbf{Block}};
\&
\&
\node{\textbf{Attributes}};
\\
\node{\begin{tabular}{@{}l@{}}backward\\exception\end{tabular}};
\&
\node[ExclusiveGateway,minimum size=8mm] (eg) {};
\node {~\,\large{\textsf{X}}};
\draw[outflow,thin] ($(eg.west)-(\outfdist,0)$) -- (eg);
\node[smallblock,right=2*\bdist of eg] (task) {\textsf{A}};
\draw[sequence,->] (eg) -- (task);
\draw[outflow,thin] (task) -- ($(task.east)+(\outfdist,0)$);
\node[IntermediateEvent,minimum size=6mm,right=0mm of task,anchor=center,yshift=-5mm,fill=white] (be) {$e$};
\node[smallblock,below right=\bdist of eg] (b) {\textsf{B}};
\draw[sequence,->,rounded corners=5pt] (be) |- (b);
\draw[sequence,->,rounded corners=5pt] (b) -| (eg);
\&
\node{
\begin{tabular}{@{}p{\legendw}@{}}
(1) Type of boundary event $e$ (error,msg,timer)\\
(2) Subprocess nested block $\textsf{A}$\\
(3) Arbitrary nested block $\textsf{B}$
\end{tabular}
};
\\
\node{\begin{tabular}{@{}l@{}}forward\\exception\end{tabular}};
\&
\node[smallblock] (a) {\textsf{A}};
\draw[outflow,thin] ($(a.west)-(\outfdist,0)$) -- (a);
\node[smallblock,right=\bdist of a] (b1) {\textsf{B}$_1$};
\node[smallblock,below=1mm of b1] (b2) {\textsf{B}$_2$};
\node[ExclusiveGateway,minimum size=8mm,right=\bdist of b1] (eg) {};
\draw[outflow,thin] (eg) -- ($(eg.east)+(\outfdist,0)$);
\node[right=\bdist of b1] {~\,\large{\textsf{X}}};
\draw[sequence,->] (a) -- (b1);
\draw[sequence,->] (b1) -- (eg);
\node[IntermediateEvent,minimum size=6mm,below=0mm of a,anchor=center,fill=white,yshift=-1mm] (be) {$e$};
\draw[sequence,->,rounded corners=5pt] (be) |- (b2);
\draw[sequence,->,rounded corners=5pt] (b2) -| (eg);
\&
\node{
\begin{tabular}{@{}p{\legendw}@{}}
(1) Type of boundary event $e$ (error,msg,timer)\\
(2) Subprocess nested block $\textsf{A}$\\
(3) Arbitrary nested blocks $\textsf{B}_1$ and $\textsf{B}_2$
\end{tabular}
};
\\
\node{\begin{tabular}{@{}l@{}}forward\\non-interrupting\\exception\end{tabular}};
\&
\node[smallblock] (a) {\textsf{A}};
\draw[outflow,thin] ($(a.west)-(\outfdist,0)$) -- (a);
\node[smallblock,right=\bdist of a] (b1) {\textsf{B}$_1$};
\node[smallblock,below=1mm of b1] (b2) {\textsf{B}$_2$};
\node[ExclusiveGateway,minimum size=8mm,right=\bdist of b1] (eg) {};
\draw[outflow,thin] (eg) -- ($(eg.east)+(\outfdist,0)$);
\node[right=\bdist of b1] {~\,\large{\textsf{X}}};
\draw[sequence,->] (a) -- (b1);
\draw[sequence,->] (b1) -- (eg);
\node[IntermediateEvent,minimum size=6mm,below=0mm of a,anchor=center,fill=white,yshift=-1mm,densely dashed] (be) {$e$};
\draw[sequence,->,rounded corners=5pt] (be) |- (b2);
\draw[sequence,->,rounded corners=5pt] (b2) -| (eg);
\&
\node{
\begin{tabular}{@{}p{\legendw}@{}}
(1) Type of boundary event $e$ (msg,timer)\\
(2) Subprocess nested block $\textsf{A}$\\
(3) Arbitrary nested blocks $\textsf{B}_1$ and $\textsf{B}_2$
\end{tabular}
};
\\
};
\end{tikzpicture}
}

\section{Preliminaries}
\label{sec:preliminaries}

In this section we give a review of the preliminaries needed to present RASs.

We adopt the usual first-order syntactic notions of signature, term (denoted with $t_1, t_2...$), atom,
(ground) formula, and so on.  We use $\uu$ to represent a tuple
$\tup{u_1,\ldots,u_n}$.  Our signatures $\Sigma$ are multi-sorted and include
equality for every sort, which implies that variables are sorted as well.
Depending on the context, we keep the sort of a variable implicit, or we
indicate explicitly in a formula that variable $x$ has sort $S$ by employing
notation $x:S$.  The notation $t(\ux)$, $\phi(\ux)$ means that the term $t$,
the formula $\phi$ has free variables included in the tuple $\ux$.
We are concerned only with constants and function symbols $f$, each of which
has \emph{sources} $\uS$ and a \emph{target} $S'$, denoted as
$f:\uS\longrightarrow S'$.  We assume that terms and formulae are well-typed,
in the sense that the sorts of variables, constants, and function
sources/targets match.  A formula is said to be \emph{universal} (resp.,
\emph{existential}) if it has the form $\forall \ux\, (\phi(\ux))$ (resp.,
$\exists \ux\, (\phi(\ux))$), where $\phi$ is a quantifier-free
formula. Formulae with no free variables are called \emph{sentences}.

From the semantic side, we use the standard notions of a
\emph{$\Sigma$-structure} $\cM$ and of \emph{truth} of a formula in a
$\Sigma$-structure under an assignment to the free variables.
A \emph{$\Sigma$-theory} $T$ is a set of $\Sigma$-sentences; a \emph{model} of
$T$ is a $\Sigma$-structure $\cM$ where all sentences in $T$ are true.  We use
the standard notation $T\models \phi$ to say that $\phi$ is true in all models
of $T$ for every assignment to the free variables of $\phi$.  We say that
$\phi$ is \emph{$T$-satisfiable} iff there is a model $\cM$ of $T$ and an
assignment to the free variables of $\phi$ that make $\phi$ true in $\cM$.


In the following, we specify transitions of an
artifact-centric system using first-order formulae.  To obtain a more compact
representation, we make use there of definable extensions as a means for
introducing so-called \emph{case-defined functions}.  We fix a signature
$\Sigma$ and a $\Sigma$-theory $T$; a \emph{$T$-partition} is a finite set
$\kappa_1(\ux), \dots, \kappa_n(\ux)$ of quantifier-free formulae
such that $T\models \forall \ux \bigvee_{i=1}^n \kappa_i(\ux)$ and
$T\models \bigwedge_{i\not=j}\forall \ux \neg (\kappa_i(\ux)\wedge
\kappa_j(\ux))$.  Given such a $T$-partition
$\kappa_1(\ux), \dots, \kappa_n(\ux)$ together with $\Sigma$-terms
$t_1(\ux), \dots, t_n(\ux)$ (all of the same target sort), a
\emph{case-definable extension} is the $\Sigma'$-theory $T'$, where
$\Sigma'=\Sigma\cup\{F\}$, with $F$ a ``fresh'' function symbol (i.e.,
$F\not\in\Sigma$)\footnote{Arity and source/target sorts for $F$ can be
 deduced from the context (considering that everything is well-typed).}, and
$T'=T \cup\bigcup_{i=1}^n \{\forall\ux\; (\kappa_i(\ux) \to F(\ux) =
t_i(\ux))\}$.
Intuitively, $F$ represents a case-defined function, which can be reformulated
using nested if-then-else expressions and can be written as
$ F(\ux) ~:=~ \mathtt{case~of}~ \{\kappa_1(\ux):t_1;\cdots;\kappa_n(\ux):t_n\}.
$ By abuse of notation, we identify $T$ with any of its case-definable
extensions $T'$.  In fact, it is easy to produce from a $\Sigma'$-formula
$\phi'$ a $\Sigma$-formula $\phi$ equivalent to $\phi'$ in all models of $T'$:
just remove (in the appropriate order) every occurrence $F(\uv)$ of the new
symbol $F$ in an atomic formula $A$, by replacing $A$ with
$\bigvee_{i=1}^n (\kappa_i(\uv) \land A(t_i(\uv)))$.
We also exploit $\lambda$-abstractions (see, e.g., formula~\eqref{eq:trans1}
below) for a more compact (still first-order) representation of some complex
expressions, and always use them in atoms like $b = \lambda y. F(y,\uz)$ as
abbreviations of $\forall y.~b(y)=F(y,\uz)$ (where, typically, $F$ is a symbol
introduced in a case-defined extension as above).

\section{Array-based Model}


In this section we recall the definitions of the formal setting presented in~\cite{CGGMR18,CGGMR19} that is exploited in this paper as target model of our translation. This setting relies on array-based systems and provides a general framework where introducing safety verification problems for artifact-centric models called Relational Artifact Systems (RASs). Those models are verbatim of the ones presented~\cite{CGGMR18,CGGMR19}, but we prefer presenting them in detail here for safe of self-containedness.

In the following, we formally define RASs. Since they are array-based systems, we start by recalling the intuition
behind them.

In general terms, an array-based system is described using a multi-sorted
theory that contains two types of sorts, one accounting for the indexes of
arrays, and the other for the elements stored therein. Since the content of an
array changes over time, it is referred to using a \emph{function}
variable, whose interpretation in a state is that of a total function mapping
indexes to elements (so that applying the function to an index denotes the
classical \emph{read} operation for arrays). The definition of an array-based
system with array state variable $a$ always requires: a formula $I(a)$
describing the \emph{initial configuration} of the array $a$, and a formula
$\tau(a,a')$ describing a \emph{transition} that transforms the content of the
array from $a$ to $a'$. In such a setting, verifying whether the system can
reach unsafe configurations described by a formula $K(a)$ amounts to check
whether the formula
$I(a_0)\wedge \tau(a_0, a_1) \wedge \cdots \wedge \tau(a_{n-1}, a_n)\wedge
K(a_n)$ is satisfiable for some $n$.

Following the tradition of artifact-centric systems \cite{DHPV09,DaDV12,boj,DeLV16}, an array-based Relational Artifact Systems (RAS) consists of a read-only DB, a read-write working memory for artifacts (which are used in our translation for formalizing the set of case variables for every process instance and the shared evolving relations), and a finite set of actions (also called services) that inspect the relational database and the working memory, and determine the new configuration of the working memory.

\subsection{Read-only DB schemata}

We now provide a formal definition of (read-only) DB-schemas by relying on an
algebraic, functional characterization.

\begin{definition}\label{def:db}
  A \emph{DB schema} is a pair $\tup{\Sigma,T}$, where:
  \begin{inparaenum}[\itshape (i)]
  \item $\Sigma$ is a \emph{DB signature}, that is, a finite multi-sorted
    signature whose only symbols are equality, unary functions, and constants;
  \item $T$ is a \emph{DB theory}, that is, a set of universal
    $\Sigma$-sentences.
  \end{inparaenum}
\end{definition}

Next, we refer to a DB schema simply through its (DB) signature $\Sigma$ and
(DB) theory $T$, and denote by $\sorts{\Sigma}$ the set of sorts and by
$\functs{\Sigma}$ the set of functions in $\Sigma$.

\begin{remark}\label{rem:extdb}

If desired, we can freely extend DB schemas by adding arbitrary $n$-ary relation symbols to the signature $\Sigma$. For this purpose, we give the
following definition.

\begin{definition}\label{def:extdb}
  A \emph{DB extended-schema} is a pair $\tup{\Sigma,T}$, where:
  \begin{inparaenum}[\itshape (i)]
  \item $\Sigma$ is a \emph{DB extended-signature}, that is, a finite multi-sorted
    signature whose only symbols are equality, $n$-ary relations, unary functions, and constants;
  \item $T$ is a \emph{DB extended-theory}, that is, a set of universal
    $\Sigma$-sentences.
  \end{inparaenum}
\end{definition}

Since for our application we are only interested in relations with primary and foreign key dependencies (even if our implementation takes into account also the case of ``free'' relations, i.e. without key dependencies),
we restrict our focus on DB schemas, which are sufficient to capture those constraints (as explained in the following subsection).
We notice that, in case the assumptions over DB schemas that
discussed below hold for DB extended-theories, all the results presented in Section~\ref{sec:artifact} (and Theorem~\ref{thm:basic}) still hold even considering DB extended-schemas instead of DB schemas.
\end{remark}

We associate to a DB signature $\Sigma$ a characteristic graph $G(\Sigma)$
capturing the dependencies induced by functions over sorts.  Specifically,
$G(\Sigma)$ is an edge-labeled graph whose set of nodes is $\sorts{\Sigma}$,
and with a labeled edge $S \xrightarrow{f} S'$ for each $f:S\longrightarrow S'$
in $\functs{\Sigma}$.
We say that $\Sigma$ is \emph{acyclic} if $G(\Sigma)$ is so. The \emph{leaves}
of $\Sigma$ are the nodes of $G(\Sigma)$ without outgoing edges.  These
terminal sorts are divided in two subsets, respectively representing
\emph{unary relations} and \emph{value sorts}. Non-value sorts (i.e., unary
relations and non-leaf sorts) are called \emph{id sorts}, and are conceptually
used to represent (identifiers of) different kinds of objects. Value sorts,
instead, represent datatypes such as strings, numbers, clock values, etc. We
denote the set of id sorts in $\Sigma$ by $\ids{\Sigma}$, and that of value
sorts by $\vals{\Sigma}$, hence
$\sorts{\Sigma} = \ids{\Sigma}\uplus\vals{\Sigma}$.

We now consider extensional data.
\begin{definition}
  \label{def:instance}
  A \emph{DB instance} of DB schema $\tup{\Sigma,T}$ is a $\Sigma$-structure
  $\cM$ that is a model of $T$ and such that every id sort of $\Sigma$ is
  interpreted in $\cM$ on a \emph{finite} set.
\end{definition}

 What may appear as not customary in Definition~\ref{def:instance} is
the fact that value sorts can be interpreted on infinite sets. This allows us,
at once, to reconstruct the classical notion of DB instance as a finite model
(since only finitely many values can be pointed from id sorts using functions),
at the same time supplying a potentially infinite set of fresh values to be
dynamically introduced in the working memory during the evolution of RASs.

We respectively denote by $S^\cM$, $f^\cM$, and $c^\cM$ the interpretation in
$\cM$ of the sort $S$ (this is a set), of the function symbol $f$ (this is a
set-theoretic function), and of the constant $c$ (this is an element of the
interpretation of the corresponding sort).  Obviously, $f^\cM$ and $c^\cM$ must
match the sorts in $\Sigma$. E.g., if $f$ has source $S$ and target $U$, then
$f^\cM$ has domain $S^\cM$ and range $U^\cM$.

We close the formalization of DB schemas by discussing DB theories, whose role
is to encode background axioms.
We illustrate a typical background axiom, required to handle the possible
presence of \emph{undefined identifiers/values} in the different sorts. This
axiom is essential to capture artifact systems whose working memory is
initially undefined, in the style of~\cite{DeLV16,verifas}.
To specify an undefined value we add to every sort $S$ of $\Sigma$ a constant
$\nullv_S$ (written from now on, by abuse of notation, just as $\nullv$, used
also to indicate a tuple).  Then, for each function symbol $f$ of $\Sigma$, we
add the following axiom to the DB theory:
\begin{equation}
  \label{eq:null}
  \forall x~(x = \nullv \leftrightarrow f(x) = \nullv)
\end{equation}
This axiom states that the application of $f$ to the undefined value produces
an undefined value, and it is the only situation for which $f$ is undefined.

As discussed in~\cite{CGGMR18}, the theory $T$ from Definition~\ref{def:db} must satisfy few crucial model-theoretic
requirements for our approach to work: these requirements are \emph{finite model property}, \emph{decidable constraint satisfiability} and the \emph{existence of $T^*$, i.e. the model completion of $T$}. Specifically, the backward reachability procedure requires the availability of suitable quantifier elimination algorithms. However, a DB theory $T$ does not necessarily have quantifier elimination; nevertheless, it is
often possible to strengthen $T$ in a conservative way (with respect to
constraint satisfiability) and get quantifier elimination. In order to do that, in~\cite{CGGMR18} we consider the theory $T^*$ (when it exists, it is unique), and  we show that model completion turns out to be quite effective to attack the verification of dynamic systems operating over relational databases. 
In all this paper we assume that DB theories $T$ have finite model property, decidable constraint satisfiability and that admit a model completion $T^*$. Specifically, from now on we assume that $T$ consists of only Axioms~\eqref{eq:null}: in this case, all the assumptions hold.

\subsection{Working Memory}

  In array-based RASs, the working memory consist of \emph{function} variables. These variables (usually called \textit{arrays})
are supposed to model evolving relations, so-called \emph{artifact
relations} in \cite{DeLV16,verifas}.  The idea is to treat artifact relations in a uniform way as
we did for the read-only DB, where we used function symbols for representing relations with key dependencies: for the working memory, we need extra sort symbols
(as explaind in the translation section, each sort symbol corresponds to a database relation symbol) and
extra unary function symbols. 
variables.

Given a DB schema $\Sigma$, an \emph{artifact extension} of $\Sigma$ is a
signature $\ext{\Sigma}$ obtained from $\Sigma$ by adding to it some extra sort
symbols\footnote{By `signature' we always mean 'signature with equality', so
 as soon as new sorts are added, the corresponding equality predicates are
 added too.}. These new sorts (usually indicated with letters $E, F, \dots$)
are called \emph{artifact sorts} (or \emph{artifact relations} by some abuse of
terminology), while the old sorts from $\Sigma$ are called \emph{basic sorts}.  
In array-based BPMN models, artifacts and basic sorts correspond, respectively, to the index and the elements sorts mentioned in the literature on  array-based systems.
Below, given $\tup{\Sigma,T}$ and an artifact extension $\ext{\Sigma}$ of
$\Sigma$, when we speak of a $\ext{\Sigma}$-model of $T$, a DB instance of
$\tup{\ext{\Sigma},T}$, or a $\ext{\Sigma}$-model of $T^*$, we mean a
$\ext{\Sigma}$-structure $\cM$ whose reduct to $\Sigma$ respectively is a model
of $T$, a DB instance of $\tup{\Sigma,T}$, or a model of $T^*$.

An \emph{artifact setting} over $\ext{\Sigma}$ is a pair $(\ux,\ua)$ given by a
finite set $\ux$ of individual variables and a finite set $\ua$ of unary
function variables: \emph{the latter are required to have an artifact sort as
 source sort and a basic sort as target sort}. Variables in $\ux$ are called
\emph{artifact variables}, and variables in $\ua$ \emph{artifact
 components}. Given a DB instance $\cM$ of $\ext{\Sigma}$, an \emph{assignment} to an
artifact setting $(\ux, \ua)$ over $\ext{\Sigma}$ is a map $\alpha$ assigning
to every artifact variable $x_i\in \ux$ of sort $S_i$ an element
$x^\alpha\in S_i^\cM$ and to every artifact component
$a_j: E_j\longrightarrow U_j$ (with $a_j\in \ua$) a set-theoretic function
$a_j^\alpha: E_j^\cM\longrightarrow U_j^\cM$. 
In our array-based RASs, artifact components and artifact variables correspond, respectively, to \textit{arrays}
 and \textit{constant arrays} (i.e., arrays with all equal elements)
mentioned in the literature on  array-based systems. Intuitevely, an artifact setting represents
the ``working'' memory of an array-based RAS.

We can view an assignment to an artifact setting $(\ux, \ua)$ as a DB instance
\emph{extending} the DB instance $\cM$ as follows.  Let all the artifact
components in $(\ux, \ua)$ having source $E$ be
$a_{i_1}: E\longrightarrow S_1, \cdots, a_{i_n}:E\longrightarrow S_n$. Viewed
as a relation in the artifact assignment $(\cM,\alpha)$, the artifact relation
$E$ ``consists'' of the set of tuples $ \{\tup{e, a_{i_1}^\alpha[e], \dots, a_{i_n}^\alpha[e]} \mid e\in E^{\cM} \}$.
Thus each element of $E$ is formed by an ``entry'' $e\in E^\cM$ (uniquely
identifying the tuple, and called ``internal identifier'' of the tuple $ (e, a_{i_1}^\alpha[e], \dots, a_{i_n}^\alpha[e])$) 
and by ``data'' $\ua_i^\alpha(e)$ taken from the
read-only database $\cM$.  When the system evolves, the set $E^\cM$ of entries
remains fixed, whereas the components $\ua_i^\alpha(e)$ may change: typically,
we initially have $\ua_i^\alpha(e)=\nullv$, but these values are changed when
some defined values are inserted into the relation modeled by $E$; the values
are then repeatedly modified (and possibly also reset to $\nullv$, if the tuple
is removed and $e$ is re-set to point to undefined values)\footnote{In
 accordance with \textsc{mcmt} conventions, we denote
 the application of an artifact component $a$ to a term (i.e., constant or
 variable) $v$ as $a[v]$ (standard notation for arrays), instead of $a(v)$.}.
 
 \subsection{RASs}

In order to introduce verification problems in the symbolic setting of array-based systems, one first
has to specify which formulae are used to represent
\begin{compactitem}
        \item sets of states,
        \item the system initializations, and
        \item system evolution.
\end{compactitem}
To introduce relational dynamic systems we discuss the kind of formulae we
use.  In such formulae, we use notations like $\phi(\uz,\ua)$ to mean that
$\phi$ is a formula whose free individual variables are among the $\uz$ and
whose free unary function variables are among the $\ua$.  Let $(\ux,\ua)$ be an
artifact setting over $\ext{\Sigma}$, where $\ux=x_1,\dots, x_n$ are the
artifact variables and $\ua=a_1,\dots,a_m$ are the artifact components (their
source and target sorts are left implicit).

An \emph{initial formula} is a formula $\iota(\ux)$ of the
  form\footnote{Recall that $a_j =\lambda y. d_{j}$ abbreviates
   $\forall y\, a_{j}(y)=d_{j}$.}
    $\textstyle
    (\bigwedge_{i=1}^n x_i= c_i) \land
    (\bigwedge_{j=1}^m a_j =\lambda y. d_j)$,
  where $c_i$, $d_j$ are constants from $\Sigma$ (typically, $c_i$  and $d_j$ are $\nullv$).
  
A \emph{state formula} has the form
 $\exists \ue\, \phi(\ue, \ux,\ua)$,
  where $\phi$ is quantifier-free and the $\ue$ are individual variables of
  artifact sorts.
  
  A \emph{transition formula} $\hat\tau$ has the form
  \begin{equation}\label{eq:trans1}
  \textstyle
    \exists \ue\,(
      \gamma(\ue,\ux,\ua)
       \land\bigwedge_i x'_i= F_i(\ue,\ux,\ua)
       \land \bigwedge_j a'_j=\lambda y. G_j(y,\ue,\ux,\ua)
    )
  \end{equation}
  where the $\ue$ are individual variables (of \emph{both} basic and artifact
  sorts), $\gamma$ (the `guard') is quantifier-free, $\ux'$, $\ua'$ are renamed
  copies of $\ux$, $\ua$, and the $F_i$, $G_j$ (the `updates') are case-defined
  functions.ed


\begin{definition}\label{def:array-model}
  An \emph{array-based RAS}  is
  $$
    \cS ~=~\tup{\Sigma,T,\ext{\Sigma}, \ux, \ua, \iota(\ux,\ua),
     \tau(\ux,\ua,\ux',\ua')}
$$
  where:
  \begin{inparaenum}[\it (i)]
  \item $\mathcal{D}\mathcal{B}:=\tup{\Sigma,T}$ is a (read-only) DB schema,
 \item $\ext{\Sigma}$ is an artifact extension of $\Sigma$,
  \item $(\ux, \ua)$ is an artifact setting over $\ext{\Sigma}$,
  \item $\iota$ is an \emph{initial formula}, and
  \item $\tau$ is a disjunction of \emph{transition formulae} $\hat\tau$.
  \end{inparaenum}
\end{definition}

\section{Parameterized Safety via Backward Reachability in RAS.}
\label{sec:artifact}

All the result presented in this section come from~\cite{CGGMR18}, where all the proofs and the details are provided.

Given a \ras $\cS$, we say that a \emph{safety} formula for $\cS$ is a state formula
$\upsilon(\ux)$ describing undesired states of $\cS$. 
As usual in array-based systems, we say that $\cS$ is \emph{safe with
        respect to} $\upsilon$ if intuitively the system has no finite run leading from
$\iota$ to $\upsilon$.  Formally, there is no DB-instance $\cM$ of $\tup{\ext{\Sigma},T}$, no $k\geq 0$, and
no assignment in $\cM$ to the variables $\ux^0,\ua^0 \dots, \ux^k, \ua^k$ such
that the formula
\begin{equation}\label{eq:smc1}
    \iota(\ux^0, \ua^0)
    \land \tau(\ux^0,\ua^0, \ux^1, \ua^1)
    \land \cdots
    \land\tau(\ux^{k-1},\ua^{k-1}, \ux^k,\ua^{k})
    \land \upsilon(\ux^k,\ua^{k})
\end{equation}
is true in $\cM$ (here $\ux^i$, $\ua^i$ are renamed copies of $\ux$, $\ua$). The \emph{safety
        problem} for $\cS$ is the following: \emph{given a safety formula $\upsilon$
        decide whether $\cS$ is safe with respect to $\upsilon$}.
        
         \begin{wrapfigure}[12]{r}{0.45\textwidth}
 \vspace{-22pt}
 \small
\begin{algorithm}[H]
\SetKwProg{Fn}{Function}{}{end}
\Fn{$\mathsf{BReach}(\upsilon)$}{
\setcounter{AlgoLine}{0}
\ShowLn$\phi\longleftarrow \upsilon$;  $B\longleftarrow \bot$\;
\ShowLn\While{$\phi\land \neg B$ is $T$-satisfiable}{
\ShowLn\If{$\iota\land \phi$ is $T$-satisfiable.}
{\textbf{return}  $\mathsf{unsafe}$}
\setcounter{AlgoLine}{3}
\ShowLn$B\longleftarrow \phi\vee B$\;
\ShowLn$\phi\longleftarrow \mathit{Pre}(\tau, \phi)$\;
\ShowLn$\phi\longleftarrow \mathsf{QE}(T^*,\phi)$\;
}
\textbf{return} $(\mathsf{safe}, B)$;}{
\caption{Backward reachability algorithm}\label{alg1}
}
\end{algorithm}
 \end{wrapfigure}
 
 In order to assess safety of Data-aware BPMN models, we run the backward reachability procedures on \ras{s}, by exploiting the translation of Data-Aware BPMN models into the array-based relational setting presented in the previous sections.

        We shall introduce an algorithm that semi-decides safety problems for $\cS$ and then we shall examine some interesting cases where the algorithm terminates and gives a decision procedure.  
Algorithm~\ref{alg1} describes the \emph{backward reachability algorithm}  for handling the safety problem for array-based systems $\cS$ .  An integral
part of the algorithm is to compute preimages.
For that purpose, for any $\phi_1(\ux,\ux')$ and $\phi_2(\ux)$, we define
$\mathit{Pre}(\phi_1,\phi_2)$ to be the formula
$\exists \ux'(\phi_1(\ux, \ux')\land \phi_2(\ux'))$.
The \emph{preimage} of the set of states described by a state formula
$\phi(\ux)$ is the set of states described by
$\mathit{Pre}(\tau,\phi)$.

 The
subprocedure $\mathsf{QE}(T^*,\phi)$ in Line~6 applies the quantifier
elimination algorithm of $T^*$ (the model completion of $T$) to the existential formula $\phi$.
Algorithm~\ref{alg1} computes iterated preimages of $\upsilon$ and applies to
them quantifier elimination, until a fixpoint is reached or until a set
intersecting the initial states (i.e., satisfying $\iota$) is found.
\textit{Inclusion} (Line~2) and \textit{disjointness} (Line~3)
tests produce proof obligations that can be 
discharged via proof obligations to be handled by SMT solvers. 
The fixpoint is reached when the test in Line~2 returns \textit{unsat}, which means that the preimage of the set of the current states is included
in the set of states reached by the backward search so far.

We obtain the following Theorem (to understand the statement of the theorem, notice that by \emph{partial correctness} we mean that, when
                        the algorithm terminates, it gives a correct answer and by \emph{effectiveness} we
                        means that all subprocedures in the algorithm can be effectively
                        executed):

\begin{theorem}\label{thm:nonsimple}
  Backward search (cf.\ Algorithm~\ref{alg1}) is effective and partially
  correct for solving safety problems for \ras\/s. Specifically, it is sound and complete 
  for detecting unsafety.
\end{theorem}

Theorem~\ref{thm:nonsimple} shows that backward search is a semi-decision procedure: if the system is
unsafe, backward search always terminates and discovers it; if the system is
safe, the procedure can diverge (but it is still correct).
Notice that the role of quantifier elimination (Line~6 of Algorithm~\ref{alg1})
is twofold:
\begin{inparaenum}[\itshape (i)]
\item It allows to discharge the fixpoint test of Line~2;
\item it ensures termination in significant cases, namely those where
  \emph{(strongly) local formulae}, introduced in the next section, are
  involved.
\end{inparaenum}

An interesting class of \ras{s} is the one where the working memory consists \textit{only} of artifact variables (without artifact relations): we call \sas{s} such systems. For \sas{s}, the following termination result holds:

\begin{theorem}\label{thm:basic}
        Let $\tup{\Sigma,T}$ be a DB schema with $\Sigma$ acyclic.  Then, for every \sas
        $\cS=\tup{\Sigma,T,\ux,\iota,\tau}$, backward search terminates and decides safety
                problems for $\cS$ in \PSPACE in the combined size of $\ux$, $\iota$, and
                $\tau$.
\end{theorem}

We remark that acyclicity of $\Sigma$ is a strong condition and it is not needed in general, and that Theorem~\ref{thm:basic} holds also for DB extended-schemas (so, even adding ``free relations'' to the DB signatures). In fact, analyzing the proof of Theorem~\ref{thm:basic}, it is clear that the decidability of the safety problems is guaranteed when in $T$ 
there are only finitely many quantifier-free formulae in which $\ux$ occur: this happens, for instance, in case $T$ has a purely relational signature or, more generally, even when $T$ is a generic first-order theory (and not just a DB (extended)-schema) that is \emph{locally finite}\footnote{We say that 
$T$ is locally finite iff for every finite tuple of variables $\ux$ there are only finitely many non $T$-equivalent atoms 
$A(\ux)$ involving only the variables $\ux$.}.

\section{Termination with local transitions}
\label{sec:termin}

We briefly recall the notion of \emph{locality} and \emph{strong locality} of transitions as presented in~\cite{CGGMR18}. All the following notions (and the following theorem) are presented in~\cite{CGGMR18}).
 
Consider an acyclic signature $\Sigma$, a DB theory $T$ and an artifact setting $(\ux,\ua)$ over an artifact
extension $\ext{\Sigma}$ of $\Sigma$.
We call a state formula \emph{local} if it is a disjunction of the formulae
\begin{equation}\label{eq:local}
  \textstyle
  \exists e_1\cdots \exists e_k\, ( \delta(e_1,\dots, e_k)  \land
    \bigwedge_{i=1}^k \phi_i(e_i,\ux,\ua)),
\end{equation}
and \emph{strongly local} if it is a disjunction of the formulae
\begin{equation}\label{eq:localstrong}
  \textstyle
  \exists e_1\cdots\exists e_k\, ( \delta(e_1, \dots, e_k) \land
  \psi(\ux) \land \bigwedge_{i=1}^k \phi_i(e_i, \ua)).
\end{equation}
In~\eqref{eq:local} and~\eqref{eq:localstrong}, $\delta$ is a conjunction of
variable equalities and inequalities, $\phi_i$, $\psi$ are quantifier-free, and
$e_1,\ldots,e_k$ are individual variables varying over artifact sorts.
The key limitation of local state formulae is that they cannot
compare entries from different tuples of artifact relations:
each $\phi_i$ in~\eqref{eq:local} and~\eqref{eq:localstrong} can contain only
the existentially quantified variable $e_i$.

A transition formula $\hat\tau$ is \emph{local} (resp., \emph{strongly local})
if whenever a formula $\phi$ is local (resp., strongly local), so is
$\mathit{Pre}(\hat\tau,\phi)$ (modulo the axioms of $T^*$). Examples of
(strongly) local $\hat\tau$ are discussed in Appendix F in~\cite{CGGMR18}.

\begin{theorem}\label{thm:term1}
  If $\Sigma$ is acyclic, backward search (cf.\ Algorithm~\ref{alg1})
  terminates when applied to a local safety formula in a \ras whose $\tau$ is a
  disjunction of local transition formulae.
\end{theorem}

\section{Translation of Data-aware BPMN models into array-based Systems}

In this section, we define the translation of a Data-aware BPMN model into array-based RAS{s}.

\subsection{Translation of the Data Schema}

\subsubsection{Catalogue.}

We now clarify how  the classical, relational database $Cat$ can be actually translated to the algebraic, functional characterization of symbolic DB schemata and
instances.  
 To technically explain the correspondence, we
adopt the \emph{named perspective}, where each relation schema is defined by a
signature containing a \emph{relation name} and a set of \emph{typed attribute
 names}. 
 
 First of all, take the set $\S$ of sorts of a \dab as set of basic sorts for the translated DB schema that we want to define. 
 Let $Cat$ be a catalogue as defined in Section~\ref{sec:data-schema}. For every $n+1$-ary relation  $R$ in $Cat$, 
 every attribute $A$ of $R$ is defined over a corresponding basic sort $S_A$ in $\S$. Since each relation $R(id,\vec{A})$ in $Cat$ must have a unary primary key as its first attribute $R.id$,
  we define a mechanism to correctly reference other attributes $\vec{A}$ in $R$ by means of unary functions in the DB signature $f_{R, A_i}: S_{R.id}\to S_{A_i}$ (where $i=1,...,n$ and $\vec{A}=(A_1,...,A_n)$):  
 $S_{R.id}$ is set as the id sort of $R$ and, in the corresponding DB instance $\cM$, $f^{\cM}_{R,A_i}$ maps, for every tuple $\ux$ in $R^{\cM}$, its identifier element (i.e., the first component of $\ux$) to
 the unique element in the tuple corresponding to the attribute $A_i$. If $A_i$ is an id sort of some other relation $R'$,  $f_{R,A_i}$ represents the foreign key referencing to $R'$.
 
 Conversely, starting from a symbolic DB schema,  we see how Definition~\ref{def:db} naturally corresponds to the definition of
relational database schema equipped with single-attribute \emph{primary keys}
and \emph{foreign keys} 

 Let $\tup{\Sigma,T}$ be a DB schema. Each id sort $S \in \ids{\Sigma}$
corresponds to a dedicated relation $R_S$ with the following attributes:
\begin{inparaenum}[\itshape (i)]
\item one identifier attribute $id_S$ with type $S$;
\item one dedicated attribute $a_f$ with type $S'$ for every function symbol
  $f \in \functs{\Sigma}$ of the form $f: S \longrightarrow S'$.
\end{inparaenum}

The fact that $R_S$ is built starting from functions in $\Sigma$ naturally
induces different database dependencies in $R_S$. In particular, for each
non-id attribute $a_f$ of $R_S$, we get a \emph{functional dependency} from
$id_S$ to $a_f$; altogether, such dependencies in turn witness that
$\mathit{id}_S$ is the \emph{(primary) key} of $R_S$. In addition, for each
non-id attribute $a_f$ of $R_S$ whose corresponding function symbol $f$ has id
sort $S'$ as image, we get an \emph{inclusion dependency} from $a_f$ to the id
attribute $id_{S'}$ of $R_{S'}$; this captures that $a_f$ is a \emph{foreign
 key} referencing $R_{S'}$.
 
 Given a DB instance $\M$ of $\tup{\Sigma,T}$, its corresponding
\emph{relational instance} $\I$ is the minimal set satisfying the following
property: for every id sort $S \in \ids{\Sigma}$, let $f_1,\ldots,f_n$ be all
functions in $\Sigma$ with domain $S$; then, for every identifier
$\constant{o} \in S^\M$, $\I$ contains a \emph{labeled fact} of the form
$R_S(\lentry{id_S}{\constant{o}^\M}, \lentry{a_{f_1}}{f_1^\M(\constant{o}^\M)},
\ldots, \lentry{a_{f_n}}{f_n^\M(\constant{o}^\M)})$.
With this interpretation, the \emph{active domain of $\I$} is the  set
\[
    \bigcup_{S \in \ids{\Sigma}} (S^\M \setminus \set{\nullv^\M}) 
    \cup
    \left\{
      \constant{v} \in \bigcup_{V \in \vals{\Sigma}} V^\M ~\left|~
        \begin{array}[c]{@{}l@{}}
          \constant{v}\neq\nullv^\M \text{ and there exist } f \in \functs{\Sigma}\\ \text{and }
          \constant{o}\in \domain{f^\M} \text{ s.t.~} f^\M(\constant{o}) =
          \constant{v}
        \end{array}
      \right.
      \!
    \right\}
\]
consisting of all (proper) identifiers assigned by $\M$ to id sorts, as well as
all values obtained in $\M$ via the application of some function. Since such
values are necessarily \emph{finitely many}, one may wonder why in
Definition~\ref{def:instance} we allow for interpreting value sorts over
infinite sets. The reason is that, in our framework, an evolving artifact
system may use such infinite provision to inject and manipulate new values into
the working memory. From the definition of active domain above, exploiting Axioms~\eqref{eq:null} we get that the membership of a tuple
 $(x_0,\dots, x_n)$ to a generic $n+1$-ary relation $R_S$  with key dependencies
(corresponding to an id sort $S$)  can be expressed in our setting by using just unary function symbols and equality:

 \begin{equation}\label{eq:relations}
  R_S(x_0, \dots, x_n) \mbox{ iff } x_0 \neq \nullv \land x_1=f_1(x_0) \land \dots \land x_n = f_n(x_0) 
  \end{equation}
  
  Hence, the representation of negated atoms is the one that directly follows from negating~\eqref{eq:relations}:

 \begin{equation}\label{eq:not-relations}
 \neg R_S(x_0,\dots,x_n) \mbox{ iff } x_0 = \nullv \lor x_1 \neq f1(x_0) \lor \dots \lor x_n \neq f_n(x_0)
 \end{equation}

Formula~\eqref{eq:relations} exactly summarizes and explicitly shows the equivalence between symbolic DB schemata and relational Catalogues. Thus, in the following we will make use of relational Catalogues or DB schemata interchangeably.

This relational interpretation of DB schemas exactly reconstructs the
requirements posed by~\cite{DeLV16,verifas} on the schema of the
\emph{read-only} database:
\begin{inparaenum}[\itshape (i)]
\item each relation schema has a single-attribute primary key;
\item attributes are typed;
\item attributes may be foreign keys referencing other relation schemas;
\item the primary keys of different relation schemas are pairwise disjoint.
\end{inparaenum}

We stress that all such requirements are natively captured in our functional
definition of a DB signature, and do not need to be formulated as axioms in the
DB theory. The DB theory is used to express additional constraints,  like that
in Axioms~\eqref{eq:null} Notice that, in order to translate Data-aware BPMN models into the 
array-based setting, we just need to consider the DB theory with Axioms~\eqref{eq:null} only: this is because the empty DB theory itself 
is able to capture the requirements of Section~\ref{sec:data-schema}, and Axioms~\eqref{eq:null} are needed to handle the undefined values in a correct way, that is every function symbol $f$ maps the undefined value of one sort to the undefined value of another one, and it is the only case when $f$ is undefined.

\subsubsection{Repository and Case Variables.}

In the following, we denote an artifact assignment and a DB instance by using $\alpha$ and $\cM$ respectively.

Consider the set of process instances $\PI$  of a data-aware BPMN model. We associate to $\PI$ a fresh sort symbol  $PI_{index}$. We call an artifact component with source $PI_{index}$``\emph{case variables} artifact component'' and we say that  all the case variables artifact components  form ``the \emph{case variables} artifact relation''. Intuitively, every tuple with first component $i\in PI_{index}$ of this artifact relation is used to formalize the values of
the case variables for the specific process instance represented by the element  $i$ that is the \emph{internal identifier} of the tuple.  We associate to every case variable $v\in V_C$ with sort $S$ an artifact component $a_v$ with $PI_{index}$ as its source sort and $S$ as its target sort. Intuitively, we are associating to every case variable an array whose locations are indexed by the process instances from $PI_{index}$ and whose components contain a value from the interpretation of the sort $S$ in $Cat$. For every process instance $i\in PI_{index}^{\cM}$, the tuple $(i, a_{v_1}^\alpha[i], \dots, a_{v_n}^\alpha[i])$ denotes the set of case variables for the process instance $i$. All the ``the \emph{case variables} artifact relation'' are usually initialized with undefined values.

Then, we associate to every relation $R:=\tup{R_1,...,R_m}$ in $Repo$ an artifact sort symbol $R_{index}$ different from $PI_{index}$ and $m$ artifact components $a_{R_1},...,a_{R_m}$ with $R_{index}$ as their common source and respectively the sorts of $R_1$,...$R_m$ as their target: sometimes, we denote the tuple $(a_{R_1},...,a_{R_m})$ by writing $a_R$. Intuitively, given a tuple $(z_1^{\cM},...,z_m^{\cM})\in R^{\cM}$, we associate to it an element $e\in R_{index}^{\cM}$ and the tuple $(e, a_{R_1}^\alpha[e], \dots, a_{R_m}^\alpha[e])$, where $a_{R_1}^\alpha[e]=z_1^{\cM}, \dots, a_{R_m}^\alpha[e]=z_m^{\cM}$: the element $e$ is the ``internal identifier'' of that tuple in $R^{\cM}$. Artifact relations that have a sort $E$ different from $PI_{index}$  as their artifact sort are called ``shared (or repository) artifact relations''. All the ``shared artifact relations'' are usually initialized with undefined values.

 \subsection{Translation of Update Logic and Process Schema}\label{subsec:trad}

We inductively translate into the array-based setting the blocks from our data-aware BPMN models that are used to construct workflows. We associate to the lifecycle status of every block $B$ a \emph{case variable artifact component} ``$lifecycleStateB$'' with $PI_{index}$ as their artifact sort. These function variables constitute the \emph{control variables} of the translated workflows. For every DAB $\cS$, we define $\mathcal{B}:= \{lifecycleB \Vert B\in Blocks\}$, where $Blocks$ is the set of all the  blocks that form the Process Component of $\cS$.

The first block that should be translated is the block Task. Since it involves preconditions and update, we preliminarily discuss how to translate them into the array-based setting.

Given a guard $q(\vec{x}) \leftarrow Q$ over $\D$ as defined in Section~\ref{sec:data-schema}, for the purpose of our translation we can assume that $Q$, instead of a disjunction of conjunctive queries with filters, only consists of a \emph{conjunction of atoms or negated atoms (i.e., cubes)}, where every atom is like in Section~\ref{sec:data-schema}. In fact, we can first put $Q$ in disjunctive normal form: so, $Q$ is equivalent to a disjunction of cubes. Then, it can be easily seen that, since existential quantifiers commute with disjunctions, it is always possibile to preprocess a precondition of a task that is  a disjunction of cubes so as to split that task into  new tasks  with preconditions that are cubes (every disjunct is the precondition of one of the new tasks). The resulting \dab is equivalent to the original one.

We say that an \emph{extended guard} is a cube $Q$ whose variables  $v$ from $\dcvars{\D}\cup \mathcal{B}$ are substituted with terms $v[I]$, where $v$ are function variables that keep the same name of $v$ and $I$ is a process instance in $\PI$. Analogously, we define \emph{extended update for a task} a formula of type ``update'' as defined in Definition~\ref{sec:update-logic}  whose variables  $v$ from $\dcvars{\D}\cup \mathcal{B}$ are substituted with terms $v[I]$, where $v$ are function variables that keep the same name of $v$ and $I$ is a process instance in $\PI$. In general, given a formulae $\phi$, we call $\phi_{ext(I)}$ the formula obtained from $\phi$ by substituting every variable $v$ from $\dcvars{\D}\cup \mathcal{B}$ in $\phi$ with $v[I]$, where $I\in \PI$.


Given an extended formulae, the translation of the query language works as follows: a variable $v\in \dcvars{\D}\cup \mathcal{B}$ is associated to a \emph{case variable artifact component} $v$ (that keeps the same name) in the array-based setting, and every term $v[I]$ (with $I\in \PI$) of $v$ is associated to the corresponding function application (read-operation) $v[i]$ (with $i\in PI_{index}$) in the array-based setting.

 We give the formal translation of the semantics of data-aware BPMN models by introducing \emph{rule-based transitions} that fit the format of transition formulae in array-based RASs. In fact, for the purpose of this paper, we can simplify the form of  transition formulae from RASs by focusing on rule-based formulae of the following form:

\begin{lstlisting}[escapeinside={(*}{*)}]
rule Transition  = 
	if  Guard
	then	Update
\end{lstlisting}

when ``Guard'' is an \emph{extended guard} and ``Update'' is an \emph{extended update for a task}.

 Formally, it can be easily seen that the previous rule-based formulae can be translated into transition formulae of RASs as follows. First of all, given an extended guard $G$ and an extended update $U$, we rewrite $G$ and $U$ into the array-based setting as follows (when we say ``add to $G$ (or to $U$) the sub-formula $\phi$'', it means that we conjunct the sub-formula $\phi$ with $G$ (or $U$)):
\begin{enumerate}
\item If U is the constructor  $\INSERT\ \vec{u}\ \INTO\ R\ \SETTING\ \vec{x}=\vec{v}$ for $R\in Repo$, and the required semantics is the set-theoretic one, substitute it in $U$ with the sub-formula:
$$\ua_{R}':=\forall j (\mbox{ if } j=e_{ins} \mbox{ then }\vec{u} \mbox{ else }(\mbox{if } \ua_{R}[j]=\vec{u} \mbox{ then }\nullv \mbox{ else } \ua_{R}[j]))\land \vec{x}':=\vec{v}$$
 where $\ua_R:=(a_{R_1},...,a_{R_n})$ are the artifact components of the shared artifact relation $R_{index}$ and $e_{ins}$ is in $R_{index}$, and add to $G$ the sub-formula $(a_{R_1}[e_{ins}]=\nullv\land...\land a_{R_n}[e_{ins}]=\nullv)$;
\item If U is the constructor $\INSERT\ \vec{u}\ \INTO\ R\ \SETTING\ \vec{x} = \vec{v}$ for $R\in Repo$,  and the required semantics is the multiset-theoretic one, substitute it in $U$ with the sub-formula $\ua_{R}'[e_{ins}]:=\vec{u}\land \vec{x}':=\vec{v}$, where $\ua_R:=(a_{R_1},...,a_{R_n})$ are the artifact components of the shared artifact relation $R_{index}$ and $e_{ins}$ is in $R_{index}$, and add to $G$ the sub-formula $(a_{R_1}[e_{ins}]=\nullv\land...\land a_{R_n}[e_{ins}]=\nullv)$;
\item If U is the constructor $\MOVE \ \vec{u}\ \FROM\ R\ \SETTING\ \vec{x}=  \vec{v}$ for $R\in Repo$, substitute it in $U$ with the sub-formula $\ua_{R}'[e_{del}]:=\nullv\land \vec{x}':=\vec{v}$, where $\ua_{R}:=(a_{R_1},...,a_{R_n})$ are artifact components of the shared artifact relation $R_{index}$ and $e_{del}$ is in $R_{index}$, and add to $G$ the subformula $\ua_{R}[e_{del}]=\vec{u}$.
\item If U is the constructor $\UPDATE\, R(\vec{v}) \,\IF\, \psi(\vec{u},\vec{v}) \,\THEN\, R(\vec{u}') \,\ELSE\, R(\vec{u}'')$ for $R\in Rep$, substitute it in $U$ with  $\ua_{R}':=\forall j (\mbox{ if } \psi(\vec{u}, \ua_{R}[j])  \mbox{ then } \vec{u}_1\mbox{ else } \vec{u}_2)$, where $\ua_{R}=(a_{R_1},...,a_{R_n})$ are artifact components of the shared artifact relation $R_{index}$, $j$ is in $R_{index}$ and $u_1,u_2$ are $u', u''$ where every occurrence of variable $v_k$ from $\vec{v}$ has been substitute with $a_{R_k}[j]$.
\item We substitute every term of the kind $v[I]$ in $G$ or in $U$, where $v\in \dcvars{\D}$, with the term $a_v[i]$, where $a_v$ is the case variables artifact component associated to $v$ and $i\in PI_{index}$;
\item We substitute every atom $R(t_1,...,t_n)$ in $G$, where $R\in Repo$, with the sub-formula $(a_{R_1}[e]=t_1\land...\land a_{R_n}[e]=t_n)$, where $a_{R_1},...,a_{R_n}$ are artifact components of the shared artifact relation $R_{index}$ and $e\in R_{index}$ is a fresh variable.
\item We substitute every atom $R(t_0,...,t_n)$ in $G$ and $U$, where $R\in Cat$, with the sub-formula $(t_0 \neq \nullv \land t_1=f_1(t_0) \land \dots \land t_n = f_n(t_0))$, where each $f_k$ ($k=1,...,n$) is the unary function $f_{R, A_k}$ associated to $R$ and its $k$-th attribute $A_k$ (here, we employ Formula~\eqref{eq:relations}).
\end{enumerate}

In Step~(4) above, we translated only the \emph{flat} ``if-then-else'' constructor, since the ``nested'' one has an analogous translation: in fact, after the keywords ``then'' or ``else'', instead of a term there will be the iterated translation of the same ``if-then-else'' constructor.

After this rewriting phase of $G$ and $U$, we obtain the following formula:

\begin{equation}\label{eq:transl}
   \exists \ue_{ins}\,\exists \ue_{del}\,\exists \ue\,\exists i\,\exists \uy\,\begin{pmatrix}
      G(v_1[i],...,v_m[i],\uy_1,\ue_{ins}, \ue_{del}, \ue) \\
       \land U(v_1[i],...,v_m[i],\uy, \ue_{ins},\ue_{del}, \ue)
    \end{pmatrix}
  \end{equation}
  
  where $ \ue_{ins}, \ue_{del}, \ue$ contains all the variables of artifact sorts that have been called $e_{ins}, e_{del}, e $ respectively during the rewriting  phase, $v_k$ are case variables artifact components associated to case variables in $\dcvars{\D}$ and $i$ is the only variable of artifact sort $PI_{index}$ introduced during the rewriting phase: notice that $i$ is different from all the  $\ue_{ins}, \ue_{del}, \ue$ variables.
Notice that we add the existential quantifier $\exists i$ in front of the transition formula and, if some variable $y\in V\setminus V_C$ occurs free in $G$ or in $U$,  we also add (avoiding redundancies) the existential quantifiers $\exists \uy$ in front of the transition formula. Then, we eliminate the quantifiers of the form $\exists y$ that bind variables of type $y$ that allow at least one definition like $y:=a_R[e]$.

Formula~\eqref{eq:transl} fits the format of Formula~\eqref{eq:trans1}. Thus, from now on it is sufficient to show that, in a rule-based transition, the formulae corresponding to ``Guard'' and ``Update'' are respectively an extended guard and an extended update. As already noticed, it is straightforward to see that preconditions and updates as presented in Section~\ref{sec:data-schema} can be transformed into extended guards and extended updates respectively.

Every block $B$ has the control variable $lifecycleStateB[i]$ for every process instance $i$ that can take at least three distinct values: ``idle'', ``enabled'', ``completed''. Blocks that have a boundary event can also have $lifecycleStateB[i]:=``error''$. We now provide the formal translation of every blocks, by exploiting rule-base transitions.\footnote{In the following, the constant ``error'' refers to finitely many different labels of type error, and each of them is linked to an exception handler: hence, every error handler is labeled using its unique error label and every constant ``error'' refers uniquely to it.}

\subsubsection{Base case: an atomic task $T$.}

In case of an atomic task, the variable $lifecycleStateT[i]$ (for every process instance $i$) takes three distinct values: ``idle'', ``enabled'', ``completed''. 
An atomic task $T$ is made up of two transitions:
\begin{itemize}
\item when $T$ is ''enabled'' for a process instance $i$, preconditions over data are evaluated and, in case they are true, $T$ can non-deterministically  update the working memory (data + control variables) and become "completed" for $i$.
\end{itemize}

We can express formally the lifecycle of an atomic task $T$ as follows:

\begin{lstlisting}[escapeinside={(*}{*)}]
rule (*$T_{1}$*) = 
	if  (*$lifecycleStateT[I]=\text{enabled}$*) 
	    PRECONDITION ON DATA
	then	(*$lifecycleStateT'[I]=\text{completed}$*)
	        UPDATES OVER THE WORKING MEMORY
\end{lstlisting}

Since ``PRECONDITION ON DATA'' and ``UPDATES OVER THE WORKING MEMORY'' are the corresponding extended versions of a guard and an update of a data-aware BPMN model respectively, and since they are conjuncted with formulae of the kind $lifecycleStateT[I]=\text{\emph{constant}}$, the previous transitions fit the format of~\eqref{eq:transl}.

In the following paragraphs, it can be easily seen that all the transitions fit the format of~\eqref{eq:transl}: specifically,  the formulae $\phi, \phi_1, \phi_2$ that appear in some Guards, are \emph{conditions} in the sense of Section~\ref{sec:data-schema} , hence the claim is true: all the other cases are straightforward.

\begin{remark} A task $T$ could be also formalized in a non-atomic way: in this case, the variable $lifecycleStateT[i]$ for every process instance $i$ takes four distinct values: ``idle'', ``enabled'', ``active'', ``completed''. 
An atomic task $T$ is made up of two transitions:
\begin{itemize}
\item when $T$ is ''enabled'' for a process instance $i$, preconditions over data are evaluated and, in case they are true, $T$ can non deterministically become ``active'' for $i$;
\item when $T$ is ``active'' for a process instance $i$, $T$ becomes "completed" for $i$ and the updates over the working memory (data + control variables) are performed.
\end{itemize}

Formally, we have:

\begin{lstlisting}[escapeinside={(*}{*)}]
rule (*$T_{1}$*) = 
	if  (*$lifecycleStateT[I]=\text{enabled}$*) 
	    PRECONDITION
	then	(*$lifecycleStateT'[I]=\text{active}$*)
\end{lstlisting}

\begin{lstlisting}[escapeinside={(*}{*)}]
rule (*$T_{2}$*) = 
	if  (*$lifecycleStateT[I]=\text{active}$*) 
	then	(*$lifecycleStateT'[I]=\text{completed}$*)
	        UPDATES OVER THE WORKING MEMORY
\end{lstlisting}
\end{remark}

\subsubsection{Base case: an event $E$.}

For every event $E$, the variable $lifecycleStateT[i]$ for every process instance $i$ takes three distinct values: ``idle'', ``enabled'', ``completed''. 
An event $E$ is made up of one transition: when $T$ is ''enabled'' for a process instance $i$, $T$ can non-deterministically become "completed" for $i$ and the updates over the working memory (data + control variables) are performed.

We can express formally the lifecycle of an event $E$ as follows:

\begin{lstlisting}[escapeinside={(*}{*)}]
rule (*$T_{1}$*) = 
	if  (*$lifecycleStateE[I]=\text{enabled}$*) 
	then	(*$lifecycleStateE'[I]=\text{completed}$*)
	        UPDATES OVER THE WORKING MEMORY
\end{lstlisting}

By using an argument similar to the previous one, we conclude that the transitions translating the behavior of an event $E$ fit the format of~\eqref{eq:transl}.

In the following, we give the translation of the \dab{s} blocks. Whenever a block $B$ has some sub-components $B_i$ (that are still blocks), we assume that they are defined by inductive hypothesis. For blocks, we use the label ``waiting'' for denoting that it is ``active''.

\inlinetitle{SEQF: sequence flow}

\vspace{2mm}

\begin{lstlisting}[escapeinside={(*}{*)}]
rule (*$T_{1}$*) = 
	if  (*$lifecycleStateB[I]=\text{enabled}$*) 
	then	(*$lifecycleStateB_1'[I]=\text{enabled}$*)
	         (*$lifecycleStateB'[I]=\text{waiting}$*)
\end{lstlisting}

\begin{lstlisting}[escapeinside={(*}{*)}]
rule (*$T_{2}$*) = 
	if  (*$lifecycleStateB_1[I]=\text{completed}$*) 
	then	(*$lifecycleStateB_1'[I]=\text{idle}$*)
		(*$lifecycleStateB_2'[I]=\text{enabled}$*)
\end{lstlisting}

\begin{lstlisting}[escapeinside={(*}{*)}]
rule (*$T_{3}$*) = 
	if  (*$lifecycleStateB_2[I]=\text{completed}$*) 
	then	(*$lifecycleStateB_2'[I]=\text{idle}$*)
		(*$lifecycleStateB'[I]=\text{completed}$*)
\end{lstlisting}

\inlinetitle{PAR: parallel block  $B$}

\vspace{2mm}

\begin{lstlisting}[escapeinside={(*}{*)}]
rule (*$T_{1}$*) = 
	if  (*$lifecycleStateB[I]=\text{enabled}$*) 
	then	(*$lifecycleStateB_1'[I]=\text{enabled}$*)
		(*$lifecycleStateB_2'[I]=\text{enabled}$*)
		(*$lifecycleStateB'[I]=\text{waiting}$*)
\end{lstlisting}

\begin{lstlisting}[escapeinside={(*}{*)}]
rule (*$T_{2}$*) = 
	if  (*$lifecycleStateB_1[I]=\text{completed}$*) 
	    (*$lifecycleStateB_2[I]=\text{completed}$*) 
	then	(*$lifecycleStateB_1'[I]=\text{idle}$*)
		(*$lifecycleStateB_2'[I]=\text{idle}$*)
		(*$lifecycleStateB'[I]=\text{completed}$*)
\end{lstlisting}

\inlinetitle{OR:  conditional inclusive block  $B$}

\vspace{2mm}

\begin{lstlisting}[escapeinside={(*}{*)}]
rule (*$T_{1}$*) = 
	if  (*$lifecycleStateB[I]=\text{enabled}$*) 
	(*$\phi_1\land (\neg \phi_2)$*) 
	then	(*$lifecycleStateB_1'[I]=\text{enabled}$*)
		(*$lifecycleStateB'[I]=\text{waiting1}$*)
\end{lstlisting}

\begin{lstlisting}[escapeinside={(*}{*)}]
rule (*$T_{2}$*) = 
	if  (*$lifecycleStateB[I]=\text{enabled}$*) 
	    (*$\phi_2\land (\neg \phi_1)$*) 
	then	(*$lifecycleStateB_2'[I]=\text{enabled}$*)
	        (*$lifecycleStateB'[I]=\text{waiting}1$*)
\end{lstlisting}

\begin{lstlisting}[escapeinside={(*}{*)}]
rule (*$T_{3}$*) = 
	if  (*$lifecycleStateB[I]=\text{enabled}$*) 
	    (*$\phi_1\land  \phi_2$*) 
	then	(*$lifecycleStateB_1'[I]=\text{enabled}$*)
	        (*$lifecycleStateB_2'[I]=\text{enabled}$*)
	         (*$lifecycleStateB'[I]=\text{waiting}2$*) 
\end{lstlisting}

\begin{lstlisting}[escapeinside={(*}{*)}]
rule (*$T_{4}$*) = 
	if  (*$lifecycleStateB_1[I]=\text{completed}$*) 
	    (*$lifecycleStateB[I]=\text{waiting}1$*) 
	then	(*$lifecycleStateB_1'[I]=\text{idle}$*)
		(*$lifecycleStateB'[I]=\text{completed}$*)
\end{lstlisting}

\begin{lstlisting}[escapeinside={(*}{*)}]
rule (*$T_{5}$*) = 
	if  (*$lifecycleStateB[I]=\text{waiting}1$*) 
	    (*$lifecycleStateB_2[I]=\text{completed}$*) 
	then	(*$lifecycleStateB_2'[I]=\text{idle}$*)
		(*$lifecycleStateB'[I]=\text{completed}$*)
\end{lstlisting}

\begin{lstlisting}[escapeinside={(*}{*)}]
rule (*$T_{6}$*) = 
	if  (*$lifecycleStateB_1[I]=\text{completed}$*) 
	    (*$lifecycleStateB_2[I]=\text{completed}$*) 
	then	(*$lifecycleStateB_1'[I]=\text{idle}$*)
		(*$lifecycleStateB_2'[I]=\text{idle}$*)
		(*$lifecycleStateB'[I]=\text{completed}$*)
\end{lstlisting}

\inlinetitle{CHOICE:  conditional exclusive block with choice $B$}

\vspace{2mm}

\begin{lstlisting}[escapeinside={(*}{*)}]
rule (*$T_{1}$*) = 
	if  (*$lifecycleStateB[I]=\text{enabled}$*) 
	    (*$\phi$*) 
	then	(*$lifecycleStateB_1'[I]=\text{enabled}$*)
	         (*$lifecycleStateB'[I]=\text{waiting}$*) 
\end{lstlisting}

\begin{lstlisting}[escapeinside={(*}{*)}]
rule (*$T_{2}$*) = 
	if  (*$lifecycleStateB[I]=\text{enabled}$*) 
	    (*$\neg\phi$*) 
	then	(*$lifecycleStateB_2'[I]=\text{enabled}$*)
	        (*$lifecycleStateB'[I]=\text{waiting}$*)
\end{lstlisting}

\begin{lstlisting}[escapeinside={(*}{*)}]
rule (*$T_{3}$*) = 
	if  (*$lifecycleStateB_1[I]=\text{completed}$*) 
	then	(*$lifecycleStateB_1'[I]=\text{idle}$*)
		(*$lifecycleStateB'[I]=\text{completed}$*)
\end{lstlisting}

\begin{lstlisting}[escapeinside={(*}{*)}]
rule (*$T_{4}$*) = 
	if (*$lifecycleStateB_2[I]=\text{completed}$*) 
	then	(*$lifecycleStateB_2'[I]=\text{idle}$*)
		(*$lifecycleStateB'[I]=\text{completed}$*)
\end{lstlisting}

\inlinetitle{DEF-CHOICE:  conditional exclusive block with deferred choice $B$}

\vspace{2mm}

\begin{lstlisting}[escapeinside={(*}{*)}]
rule (*$T_{1}$*) = 
	if  (*$lifecycleStateB[I]=\text{enabled}$*) 
	then	(*$lifecycleStateB_1'[I]=\text{enabled}$*)
		(*$lifecycleStateB'[I]=\text{waiting}$*)
\end{lstlisting}

\begin{lstlisting}[escapeinside={(*}{*)}]
rule (*$T_{2}$*) = 
	if  (*$lifecycleStateB[I]=\text{enabled}$*)  
	then	(*$lifecycleStateB_2'[I]=\text{enabled}$*)
	        (*$lifecycleStateB'[I]=\text{waiting}$*)
\end{lstlisting}

\begin{lstlisting}[escapeinside={(*}{*)}]
rule (*$T_{3}$*) = 
	if  (*$lifecycleStateB_1[I]=\text{completed}$*) 
	then	(*$lifecycleStateB_1'[I]=\text{idle}$*)
		(*$lifecycleStateB'[I]=\text{completed}$*)
\end{lstlisting}

\begin{lstlisting}[escapeinside={(*}{*)}]
rule (*$T_{4}$*) = 
	if (*$lifecycleStateB_2[I]=\text{completed}$*) 
	then	(*$lifecycleStateB_2'[I]=\text{idle}$*)
		(*$lifecycleStateB'[I]=\text{completed}$*)
\end{lstlisting}

\inlinetitle{LOOP: a loop block $B$}

\vspace{2mm}

\begin{lstlisting}[escapeinside={(*}{*)}]
rule (*$T_{1}$*) = 
	if  (*$lifecycleStateB[I]=\text{enabled}$*) 
	then	(*$lifecycleStateB_1'[I]=\text{enabled}$*)
	        (*$lifecycleStateB'[I]=\text{waiting}$*)
\end{lstlisting}

\begin{lstlisting}[escapeinside={(*}{*)}]
rule (*$T_{2}$*) = 
	if  (*$lifecycleStateB_1[I]=\text{completed}\land \phi$*)
	then	(*$lifecycleStateB_1'[I]=\text{idle}$*)
		(*$lifecycleStateB_2'[I]=\text{enabled}$*)
\end{lstlisting}

\begin{lstlisting}[escapeinside={(*}{*)}]
rule (*$T_{3}$*) = 
	if   (*$lifecycleStateB_2[I]=\text{completed}$*)  
	then	(*$lifecycleStateB_1'[I]=\text{enabled}$*)
		(*$lifecycleStateB_2'[I]=\text{idle}$*)
\end{lstlisting}

\begin{lstlisting}[escapeinside={(*}{*)}]
rule (*$T_{4}$*) = 
	if  (*$lifecycleStateB_1[I]=\text{completed}\land (\neg\phi)$*) 
	then	(*$lifecycleStateB_1'[I]=\text{idle}$*)
		(*$lifecycleStateB'[I]=\text{completed}$*)
\end{lstlisting}

\inlinetitle{PROC: a process block $B$}

\vspace{2mm}

\begin{lstlisting}[escapeinside={(*}{*)}]
rule (*$T_{1}$*) = 
	if  (*$lifecycleStateB[I]=\text{enabled}$*) 
	then	(*$lifecycleStateB_1'[I]=\text{enabled}$*)
	        (*$lifecycleStateB'[I]=\text{waiting}$*) 
\end{lstlisting}

\begin{lstlisting}[escapeinside={(*}{*)}]
rule (*$T_{2}$*) = 
	if  (*$lifecycleStateB_1[I]=\text{completed}$*)
	then	(*$lifecycleStateB_1'[I]=\text{idle}$*)
		(*$lifecycleStateB'[I]=\text{completed}$*)
\end{lstlisting}


\inlinetitle{BLOCK-ERR: a block with error $B$}

\vspace{2mm}

\begin{lstlisting}[escapeinside={(*}{*)}]
rule (*$T_{1}$*) = 
	if  (*$lifecycleStateB[I]=\text{enabled}$*) 
	then	(*$lifecycleStateB_1'[I]=\text{enabled}$*)
	        (*$lifecycleStateB'[I]=\text{waiting}$*) 
\end{lstlisting}

\begin{lstlisting}[escapeinside={(*}{*)}]
rule (*$T_{2}$*) = 
	if  (*$lifecycleStateB_1[I]=\text{completed}\land \phi$*)
	then	(*$lifecycleStateB_1'[I]=\text{idle}$*)
		(*$lifecycleStateB'[I]=\text{completed}$*)
\end{lstlisting}

\begin{lstlisting}[escapeinside={(*}{*)}]
rule (*$T_{3}$*) = 
	if  (*$lifecycleStateB_1[I]=\text{completed}\land (\neg\phi)$*)
	then	(*$lifecycleStateB_1'[I]=\text{idle}$*)
		(*$lifecycleStateH'[I]=\text{idle}$*)
		(*$lifecycleStateB_{\text{bound-hand}}'[I]=\text{error}$*)
\end{lstlisting}

where the $H$s are all the sub-blocks of the block $B_{\text{bound-hand}}$ whose boundary is directly connected to the handler block for ``error''.

Alternatively, $T_3$ could also be

\begin{lstlisting}[escapeinside={(*}{*)}]
rule (*$T_{3, alt}$*) = 
	if  (*$lifecycleStateB_1[I]=\text{completed}\land (\neg\phi)$*)
	then	(*$lifecycleStateB_1'[I]=\text{idle}$*)
		(*$lifecycleStateH'[I]=\text{idle}$*)
		(*$errorB_{\text{bound-hand}}'[I]=\true$*)
\end{lstlisting}

where $errorB_{\text{bound-hand}}$ is a boolean variable linked to $B_{\text{bound-hand}}$.



\inlinetitle{BACK-EXCP: a backward exception handling block $B$}

\vspace{2mm}

Here, $A$ is a subprocess.

\begin{lstlisting}[escapeinside={(*}{*)}]
rule (*$T_{1}$*) = 
	if  (*$lifecycleStateB[I]=\text{enabled}$*)  
	then	(*$lifecycleStateA'[I]=\text{enabled}$*)
	        (*$lifecycleStateB'[I]=\text{waiting}$*) 
\end{lstlisting}

\begin{lstlisting}[escapeinside={(*}{*)}]
rule (*$T_{2}$*) = 
	if  (*$lifecycleStateA[I]=\text{completed}$*) 
	then	(*$lifecycleStateA'[I]=\text{idle}$*)
	        (*$lifecycleStateB'[I]=\text{completed}$*)
\end{lstlisting}

\begin{lstlisting}[escapeinside={(*}{*)}]
rule (*$T_{3}$*) = 
	if  (*$lifecycleStateA[I]=\text{error}$*)
	then	(*$lifecycleStateA'[I]=\text{idle}$*)
		(*$lifecycleStateB_1'[I]=\text{enabled}$*)
\end{lstlisting}

\begin{lstlisting}[escapeinside={(*}{*)}]
rule (*$T_{err1}$*) = 
	if  (*$lifecycleStateA[I]=\text{enabled}$*)
	then	(*$lifecycleStateA'[I]=\text{error}$*)
		(*$lifecycleStateH'[I]=\text{idle}$*)
\end{lstlisting}

\begin{lstlisting}[escapeinside={(*}{*)}]
rule (*$T_{err2}$*) = 
	if  (*$lifecycleStateA[I]=\text{waiting}$*)
	then	(*$lifecycleStateA'[I]=\text{error}$*)
		(*$lifecycleStateH'[I]=\text{idle}$*)
\end{lstlisting}

\begin{lstlisting}[escapeinside={(*}{*)}]
rule (*$T_{err3}$*) = 
	if  (*$lifecycleStateA[I]=\text{active}$*)
	then	(*$lifecycleStateA'[I]=\text{error}$*)
		(*$lifecycleStateH'[I]=\text{idle}$*)
\end{lstlisting}

where the $H$s are all the sub-blocks of the block $A$.

\inlinetitle{FOR-EXCP: a forward exception handling block $B$}

\vspace{2mm}

Here, $A$ is a subprocess.

\begin{lstlisting}[escapeinside={(*}{*)}]
rule (*$T_{1}$*) = 
	if  (*$lifecycleStateB[I]=\text{enabled}$*) 
	then	(*$lifecycleStateA'[I]=\text{enabled}$*)
	         (*$lifecycleStateB'[I]=\text{waiting}$*)
\end{lstlisting}

\begin{lstlisting}[escapeinside={(*}{*)}]
rule (*$T_{2}$*) = 
	if  (*$lifecycleStateA[I]=\text{completed}$*) 
	then	(*$lifecycleStateA'[I]=\text{idle}$*)
	        (*$lifecycleStateB_1'[I]=\text{enabled}$*)
\end{lstlisting}

\begin{lstlisting}[escapeinside={(*}{*)}]
rule (*$T_{3}$*) = 
	if  (*$lifecycleStateB_1[I]=\text{completed}$*) 
	then	(*$lifecycleStateB_1'[I]=\text{idle}$*)
	        (*$lifecycleStateB'[I]=\text{completed}$*)
\end{lstlisting}

\begin{lstlisting}[escapeinside={(*}{*)}]
rule (*$T_{4}$*) = 
	if  (*$lifecycleStateA[I]=\text{error}$*)
	then	(*$lifecycleStateA'[I]=\text{idle}$*)
		(*$lifecycleStateB_2'[I]=\text{enabled}$*)
\end{lstlisting}

\begin{lstlisting}[escapeinside={(*}{*)}]
rule (*$T_{5}$*) = 
	if  (*$lifecycleStateB_2[I]=\text{completed}$*)
	then	(*$lifecycleStateB_2'[I]=\text{idle}$*)
		(*$lifecycleStateB'[I]=\text{completed}$*)
\end{lstlisting}

\begin{lstlisting}[escapeinside={(*}{*)}]
rule (*$T_{err1}$*) = 
	if  (*$lifecycleStateA[I]=\text{enabled}$*)
	then	(*$lifecycleStateA'[I]=\text{error}$*)
		(*$lifecycleStateH'[I]=\text{idle}$*)
\end{lstlisting}

\begin{lstlisting}[escapeinside={(*}{*)}]
rule (*$T_{err2}$*) = 
	if  (*$lifecycleStateA[I]=\text{waiting}$*)
	then	(*$lifecycleStateA'[I]=\text{error}$*)
		(*$lifecycleStateH'[I]=\text{idle}$*)
\end{lstlisting}

\begin{lstlisting}[escapeinside={(*}{*)}]
rule (*$T_{err3}$*) = 
	if  (*$lifecycleStateA[I]=\text{active}$*)
	then	(*$lifecycleStateA'[I]=\text{error}$*)
		(*$lifecycleStateH'[I]=\text{idle}$*)
\end{lstlisting}

where the $H$s are all the sub-blocks of the block $A$.

\inlinetitle{NON-INTERR: a non-interrupting exception handling block $B$}

\vspace{2mm}

Here, we also need a boolean variable $errorA$ that changes its value when the error occurs. $A$ is a subprocess.

\begin{lstlisting}[escapeinside={(*}{*)}]
rule (*$T_{1}$*) = 
	if  (*$lifecycleStateB[I]=\text{enabled}$*) 
	then	(*$lifecycleStateA'[I]=\text{enabled}$*)
	        (*$lifecycleStateB'[I]=\text{waiting}$*) 
\end{lstlisting}

\begin{lstlisting}[escapeinside={(*}{*)}]
rule (*$T_{2}$*) = 
	if  (*$lifecycleStateA[I]=\text{completed}$*) 
	then	(*$lifecycleStateA'[I]=\text{idle}$*)
	        (*$lifecycleStateB_1'[I]=\text{enabled}$*)
\end{lstlisting}

\begin{lstlisting}[escapeinside={(*}{*)}]
rule (*$T_{3}$*) = 
	if  (*$lifecycleStateB_1[I]=\text{completed}$*) 
	    (*$errorA=\false$*)
	then	(*$lifecycleStateB_1'[I]=\text{idle}$*)
	        (*$lifecycleStateB'[I]=\text{completed}$*)
\end{lstlisting}

\begin{lstlisting}[escapeinside={(*}{*)}]
rule (*$T_{4}$*) = 
	if  (*$errorA=\true$*)
	then	(*$lifecycleStateB_2'[I]=\text{enabled}$*)
	         (*$errorA'=\true$*)
\end{lstlisting}

\begin{lstlisting}[escapeinside={(*}{*)}]
rule (*$T_{5}$*) = 
	if   (*$lifecycleStateB_1[I]=\text{completed}$*)
	     (*$lifecycleStateB_2[I]=\text{completed}$*)
	then	(*$lifecycleStateB_1'[I]=\text{idle}$*)
	         (*$lifecycleStateB_2'[I]=\text{idle}$*)
		(*$lifecycleStateB'[I]=\text{completed}$*)
\end{lstlisting}

\begin{lstlisting}[escapeinside={(*}{*)}]
rule (*$T_{err1}$*) = 
	if  (*$lifecycleStateA[I]=\text{enabled}$*)
	then	(*$lifecycleStateA'[I]=\text{error}$*)
		(*$lifecycleStateH'[I]=\text{idle}$*)
\end{lstlisting}

\begin{lstlisting}[escapeinside={(*}{*)}]
rule (*$T_{err2}$*) = 
	if  (*$lifecycleStateA[I]=\text{waiting}$*)
	then	(*$lifecycleStateA'[I]=\text{error}$*)
		(*$lifecycleStateH'[I]=\text{idle}$*)
\end{lstlisting}

\begin{lstlisting}[escapeinside={(*}{*)}]
rule (*$T_{err3}$*) = 
	if  (*$lifecycleStateA[I]=\text{active}$*)
	then	(*$lifecycleStateA'[I]=\text{error}$*)
		(*$lifecycleStateH'[I]=\text{idle}$*)
\end{lstlisting}

where the $H$s are all the sub-blocks of the block $A$.

\subsection{Alternative translation of specific blocks}

For sake of simplicity (and in order to make the translation more efficient in performance), sometimes we can freely employ the following alternative translation for specific blocks that are useful in practice.

\inlinetitle{$n$-SEQ: $n$-iterated sequence flow block $B$}

\vspace{2mm}

In case a sequence flow block $B$ is formed of $n>2$ sub-blocks $B_k$, we can make use of the following rule-based transitions: 

\begin{lstlisting}[escapeinside={(*}{*)}]
rule (*$T_{1}$*) = 
	if  (*$lifecycleStateB[I]=\text{enabled}$*) 
	then	(*$lifecycleStateB_1'[I]=\text{enabled}$*)
	         (*$lifecycleStateB'[I]=\text{waiting}$*)
\end{lstlisting}

\begin{lstlisting}[escapeinside={(*}{*)}]
rule (*$T_{2}$*) = 
	if  (*$lifecycleStateB_1[I]=\text{completed}$*) 
	then	(*$lifecycleStateB_1'[I]=\text{idle}$*)
		(*$lifecycleStateB_2'[I]=\text{enabled}$*)
\end{lstlisting}

\begin{lstlisting}[escapeinside={(*}{*)}]
rule (*$T_{3}$*) = 
	if  (*$lifecycleStateB_k[I]=\text{completed}$*) 
	then	(*$lifecycleStateB_k[I]=\text{idle}$*)
		(*$lifecycleStateB_{k+1}'[I]=\text{enabled}$*)
\end{lstlisting}

where $2\leq k < n$

\begin{lstlisting}[escapeinside={(*}{*)}]
rule (*$T_{4}$*) = 
	if  (*$lifecycleStateB_n[I]=\text{completed}$*) 
	then	(*$lifecycleStateB_n'[I]=\text{idle}$*)
		(*$lifecycleStateB'[I]=\text{completed}$*)
\end{lstlisting}

\inlinetitle{ERR\&EVENT: ``either error or event'' block}

\vspace{2mm}

In case of a event-based fork block, where the first branch has an event $B_1$ and the second one is an error event $B_2$, we have the following translation:

\begin{lstlisting}[escapeinside={(*}{*)}]
rule (*$T_{1}$*) = 
	if  (*$lifecycleStateB_1[I]=\text{enabled}$*) 
	then	(*$lifecycleStateB_1'[I]=\text{completed}$*)
		 UPDATES OVER THE WORKING MEMORY
\end{lstlisting}

\begin{lstlisting}[escapeinside={(*}{*)}]
rule (*$T_{2}$*) = 
	if  (*$lifecycleStateB_1[I]=\text{enabled}$*) 
	then    (*$lifecycleStateB_1'[I]=\text{idle}$*)
	        (*$lifecycleStateH'[I]=\text{idle}$*)
	        (*$lifecycleStateB_{\text{bound-hand}}'[I]=\text{error}$*)
	        UPDATES OVER THE WORKIN MEMORY PERFORMED BY (*$B_2$*)
\end{lstlisting}

where the $H$s are all the sub-blocks of the block $B_{\text{bound-hand}}$ whose boundary is directly connected to the handler block for ``error''.

\subsection{Translation of Reachability Queries}

The translation of a reachability query is totally analogous to the translation presented in Subsection~\ref{subsec:trad} for guards and updates, with the proviso that case variables associated to different process instances must be translated into case variables artifact component applied to different indexes from $PI_{index}$. 
More formally, given a reachability query $Q:=\bigwedge_{i \in I} G_i[i]$, where $G_i:=\bigvee_k G_k$ is a guard, we associate to every $i\in I$ and index $e_i\in PI_{index}$, we substitute every case variable $v$ in $G_i[i]$ with the term $v[i]$ (read-operation of a function variable $x$) and we apply the same rewriting policy of Subsection~\ref{subsec:trad} to every its conjunct $G_k$: thus, we obtain a quantifier-free formula of the kind:
$$\exists \ue\, \phi(\ue, \ux,\ua)$$,
  where $\phi$ is quantifier-free and the $\ue$ are individual variables of
  artifact sorts (where each $e$ is either in $PI_{index}$ or in some ``repository'' artifact relation), i.e. we get a state formula of RASs, as required.

\section{Soundness and Completeness Results}

In this section we sketch the proof of the soundness and completeness results.

\begin{remark}\label{rem:bound}
Notice that, in our translation to array-based systems, case variables of case-bounded \dab{s} can be treated as proper \emph{artifact variables} of RASs, instead of arrays. This is trivial since we do not need (undounded)
indexes (taken from some artifact sort) in case of 1-case \dab{s}, and, analogously, in case of $k$-bounded \dab{s}, it is sufficient to associate every case variable to $k$ corresponding artifact variables.
\end{remark}

Theorem~\ref{thm:case-bounded} and Theorem~\ref{thm:unrestricted} clearly follow from Theorem~\ref{thm:nonsimple} and from the translation into RASs described in the previous section.

\section{Termination Results}

Notice that  clearly $\cat$ is acyclic iff its corresponding DB schema in RASs is acyclic. Hence, a \dab is acyclic iff its translated RAS has an acyclic DB schema.

Thanks to Remark~\ref{rem:bound}, Theorem~\ref{thm:dec-case-repo-bounded} simply follows from Theorem~\ref{thm:basic} and from our translation in RASs.

It can be easily seen, by exploiting our translation into RASs, that the conditions posed over updates in Theorem~\ref{thm:dec-case-bounded} and Theorem~\ref{thm:dec-case-unbounded} exactly correspond to suitable requirements over the translated updates in RASs that guarantee strong locality. Hence, it is possible to apply Theorem~\ref{thm:term1} in order to get termination: in fact, it is clear that the translation of a separated reachability query is a strongly local formula.

We devote the following subsection to sketch the proof of strong locality of the transitions translating the restricted updates of Theorem~\ref{thm:dec-case-bounded}.

\subsection{Translated Updates that are strongly local}

We give a sketch of the proof of strong locality of transitions that are the translations into the array-based setting of ``Insert\&set rule'' and ''Delete\&set rule'' with the restriction of Section~\ref{sec:termination}: the other updates presented in Section~\ref{sec:termination} can be proved in a similar way to have a strongly local translation. All the proofs sketched in this section concerning the fact that the format of those transitions fits the definition of strong locality are similar to the ones in Appendix F of~\cite{CGGMR18}, where all the details are deeply analyzed and \emph{all} the restricted updates corresponding to the ones presented in Section~\ref{sec:termination} are proved to be strongly local transitions.
Specifically, by adopting the conventions of~\cite{CGGMR18}, we notice that the proofs of strong locality of the translations of ``Insert\&set rule'', ``Set rule", ``Delete\&set rule'' and ``Conditional update rule'' correspond and are analogous (with trivial adaptations) to the proofs of strong locality (provided in Appendix F in~\cite{CGGMR18}) of ``Insertion Updates'', `Propagation Updates'', ``Deletion Updates'' and ``Bulk Updates''. In the following, when we say that a translated array-based formula is ``repository-free'', ``over the Catalogue'' or etc., we mean, by abuse of notation, that this formula is the translation of a corresponding ``repository-free'', ``over the Catalogue'' or etc. query of a \dab.

\subsubsection{Delete\&set rule.}
We want to remove a tuple $\underline{t}:=(t_{1},...,t_{m})$ from an $m$-ary  relation $R$ of the Repository and assign the values $t_{1},...,t_{m}$ to some of the case variables (let $\ux:=\ux_1,\ux_2$, where  $\ux_{1}:=(x_{i_{1}},...,x_{i_{m}})$ are the variables where we want to transfer the tuple $\underline{t}$). This operation has to be applied only if the current case variables $\ux$ satisfy the repository-free pre-condition $\pi(\ux_1, \ux_2)$ and additional variables $\uy:=\uy_1, \uy_2$ (where the $\uy_1$ are elements from the tuple $\underline{t}$) satisfy the post-condition $\psi(\uy_1, \uy_2)$ over the Catalogue. The variables $\ux_2$ are not propagated, i.e. they are reassigned (possibly with the same values that they have before).
Let $\underline{r}:=r_1,...,r_m$ be the artifact components of $R$ in the translated array-based setting.
 Such an update can be formalized in the translated array-based formalism as follows:

\begin{equation}\label{eq:del}
\exists \uy_1\,\exists \uy_2\,\exists i\,\exists e \, \exists \ui \begin{pmatrix}
 \pi(\ux_1[i], \ux_2[i])\;\wedge\; \underline{r}[e]=\uy_1\;\wedge\;  \psi(\uy_1, \uy_2)
\; \wedge r_1[e]\neq \nullv\wedge...\\ \wedge\; r_n[e]\neq \nullv
\wedge (\ux_1^{\prime}[i]:=\underline{r}[e]\;\wedge \;\ux_2^{\prime}[i]:=\uy_2\wedge \underline{s}^{\prime}:=\underline{s}\;\wedge \\
\;\wedge\; \underline{r}^{\prime}:=\lambda j.(\mathtt{ if }\  j=e \mathtt{~ then ~\nullv ~else ~} \underline{r}[j]))\end{pmatrix}
\end{equation}
where $\underline{s}$ are the artifact components of the relations from the Repository different from $R$, and $\pi$ and $\psi$ are free-repository conjunctive queries. Notice that the $\uy_1, \uy_2$ are non deterministically produced values for the updated $\ux^{\prime}_2$. In the terminology of \cite{verifas}, notice that no case variable variable is ``propagated'' in a deletion update.

 We sketch the proof of the fact that the preimage along \eqref{eq:del} of a strongly local formula is strongly local.
 Consider a strongly local formula
 \[
   K:=\psi^{\prime}(\ux[i])\wedge\exists \underline{e} \left(
     \text{Diff}(\underline{e})  \wedge \bigwedge_{e_r \in \underline{e}}
     \phi_{e_r}( \underline{r}[e_r])\wedge\Theta\right)
   \]
 where  $\Theta$ is a formula involving the artifact components $\underline{s}$ (which are not updated) such that no $e_r$ occurs in it.
 
 Computing the preimage $Pre(\ref{eq:del},K)$, we get (with a computation analogous to the one done in Appendix F in~\cite{CGGMR18})
the disjunction of the formulae:

 \begin{itemize}
 \item $\exists e, \underline{e} \left(\begin{array}{@{}l@{}}
\text{Diff}(\underline{e},e)\;  \wedge\; \pi(\ux_1[i],\ux_2[i])\; \wedge \; \theta(\underline{r}[e])\; \wedge \; \bigwedge_{e_r \in \underline{e}} \phi_{e_r}(\underline{r}[e_r])\; \wedge\;   \Theta
\end{array}\right)
$
\item $
\exists \underline{d}\, \exists\underline{e} \left(\begin{array}{@{}l@{}}
\text{Diff}(\underline{e})\;  \wedge\; \pi(\ux_1[i],\ux_2[i])\; \wedge  \; \theta(\underline{r}[e_j]) \; \wedge\\ \wedge  \bigwedge_{ e_r \in \underline{e}, e_r\neq e_j} \phi_{e_r}(\underline{r}[e_r])\; \wedge \; \phi_{e_j}(\nullv)\;  \wedge\; \Theta
\end{array}\right)$

\end{itemize}
which is strongly local,  where $\theta$ is a quantifier-free $\Sigma$-formula ($\Sigma$ is the DB signature of the read-only DB that is translation of the Catalogue).

\subsubsection{Insert\&set rule.}

We want to insert a tuple of values $\underline{t}:=(t_{1},...,t_{m})$ from the case variables $\ux_{1}:=(x_{i_{1}},...,x_{i_{m}})$ (let $\ux:=\ux_1,\ux_2$ as above) into an $m$-ary  relation $R$ of the Repository. This operation has to be applied only if the current case variables $\ux$ and additional variables $\uy:=\uy_1, \uy_2$ satisfy the repository-free pre-condition $\pi(\ux_1, \ux_2, \uy)$.
Let $\underline{r}:=r_1,...,r_m$ be the artifact components of $R$.
 Such an update can be formalized in the translated array-based formalism as follows:
 
 \begin{equation}\label{eq:ins}
\exists \underline{d}_1,\underline{d}_2\,\exists e \begin{pmatrix}
\pi(\ux_1[i], \ux_2[i], \uy)
\; \wedge\: \underline{r}[e]=\nullv\\
\wedge \; (\ux^{\prime}[i]:=\uy\; \wedge\;  \underline{s}^{\prime}:=\underline{s}\; \wedge \\
\wedge\; \underline{r}^{\prime}:=\lambda j.(\mathtt{ if }\  j=e \mathtt{~ then ~} \ux_1[i]\mathtt{~else ~} \underline{r}[j]))\end{pmatrix}
\end{equation}
where $\underline{s}$ are the artifact components of the relations from the Repository different from $R$. Notice that $\uy$ are used to produce values for the updated case variables $\ux^{\prime}$. In the terminology of \cite{verifas}, notice that no artifact variable is propagated in a insertion update. Notice that it is allowed that some case variables are propagated (i.e., that some, or all, $x_k':=x_k$)
 
  Notice also that the following arguments remain the same even if $\underline{r}[e]=\nullv$ is replaced with a conjunction of \textit{some} literals of the form $r_j[e]=\nullv$, for some $j=1,...,m$, or even if $\underline{r}[e]=\nullv$ is replaced with a generic constraint $\chi(\underline{r}[e])$.
  
  We sketch the proof of the fact that the preimage along \eqref{eq:ins} of a strongly local formula is strongly local.
 Consider a strongly local formula

 \[
   K:=\psi^{\prime}(\ux[i])\wedge\exists \underline{e} \left(
     \text{Diff}(\underline{e})  \wedge \bigwedge_{e_r \in \underline{e}}
     \phi_{e_r}( \underline{r}[e_r])\wedge\Theta\right)
 \]
 where  $\Theta$ is a formula involving the translation of the relations $\underline{s}$ (which are not updated) from the Repository  such that no $e_r$ occurs in it.
 
 Computing the preimage $Pre(\ref{eq:ins},K)$, we get (with a computation analogous to the one done in Appendix F in~\cite{CGGMR18})
the disjunction of the formulae:

  \begin{itemize}
 \item $\exists e, \underline{e} \left(\begin{array}{@{}l@{}}
\text{Diff}(\underline{e},e)\;  \wedge\; \theta(\ux_1[i],\ux_2[i])\; \wedge \; \underline{r}[e]=\nullv\;  \wedge \; \bigwedge_{e_r \in \underline{e}} \phi_{e_r}(\underline{r}[e_r])\; \wedge \;  \Theta
\end{array}\right)
$
\item $
\exists\underline{e} \left(\begin{array}{@{}l@{}}
\text{Diff}(\underline{e}) \; \wedge\;  \theta(\ux_1[i],\ux_2[i])\; \wedge \; \phi_{e_j}(\ux_1[i])\; \wedge  \; \underline{r}[e]=\nullv  \wedge\bigwedge_{ e_r \in \underline{e}, e_r\neq e_j} \phi_{e_r}(\underline{r}[e_r])\; \wedge \; \Theta
\end{array}\right)$

\end{itemize}
 which is a strongly local formula, where $\theta$ is a quantifier-free $\Sigma$-formula ($\Sigma$ is the DB signature of the read-only DB that is translation of the Catalogue).

We remark that, in a ``Insert\&set'' update, the insertion of the same content in correspondence to different entries is allowed. \emph{If we want to avoid this kind of multiple insertions}, the update $r^{\prime}$ must be modified as follows:

\[
  \underline{r}^{\prime}:=\lambda j.\left(\begin{array}{@{}l@{}}\mathtt{ if }\
      j=e \mathtt{~ then ~} \ux_1\mathtt{~else ~}\\ \mathtt{(if~}
      \underline{r}[j]=\ux_1 \mathtt{~ then ~ \nullv ~ else~ } \underline{r}[j]
      )\end{array}\right)
      \]

which is not strongly local.

\begin{example}

\begin{figure}[t]
\centering
\resizebox{1\textwidth}{!}{
\begin{tikzpicture}[auto,x=1.4cm,y=.9cm, thick,minimum size=.8cm]

\node[draw,rectangle,rounded corners,minimum width = 12.2cm,minimum height=.9cm] (applicant) at (6.6,2.5) {Applicant} ;
\node[draw,rectangle,rounded corners,minimum width = 12.2cm,minimum height=4.7cm] (spb) at (6.6,-1.6) {} ;

\node[StartEvent] (jobe) at (1.5,0) {};
\node[lbl,below of=jobe,align = center] (jobe_name) {Job \\ posted};
\draw[sequence,->]  (jobe) -- (2.28,0);

\node[StartEvent] (se1) at (2.7,0) {};
\node[lbl,below of=se1,align = center] (se1_name) {};
\node[ExclusiveGateway,draw] (eg1) at (3.7,0) {};
\draw[sequence,->]  (se1) -- (eg1);

\node[ExclusiveGateway,draw] (eg_err) at (4.8,0) {};
\draw[sequence,->]  (eg1) -- (eg_err);

\node[MessageStartEvent] (appe) at (6,0) {};
\node[lbl,below of=appe,align = center,yshift=1mm] (appe_name) {Application\\ received};
\draw[sequence,->]  (eg_err) -- 
(appe);

\node[ErrorEndEvent,draw,minimum size=.8cm] (err) at (6,-2.2) {};
\draw[sequence,rounded corners=5pt,->] (eg_err) |- 
(err);

\node[circle,draw=black, fill=white, inner sep=0pt,minimum size=4pt] (c) at (6,2) {};
\draw[sequence,dashed,->]  (c) -- (appe);

\node[task,align=center] (evalcv) at (7.3,0) {\tname{Evaluate}\\\tname{CV}};
\draw[sequence,->]  (appe) -- (evalcv);

\node[ExclusiveGateway,draw] (eg2) at (8.6,0) {};
\draw[sequence,->]  (evalcv) -- (eg2);

\node[EndEvent] (ee1) at (10.5,0) {};
\node[lbl,below of=ee1,align = center] (ee1_name) {};
\draw[sequence,->]  (eg2) -- node[guard,xshift=-1.5mm,yshift=-3mm,align=left] {$qualif=$\\\true} (ee1);

\node[task,align=center] (evalapp) at (7.2,-3.3) {\tname{Evaluate}\\\tname{Application}};
\draw[sequence,->,rounded corners=5pt] (eg2) |- node[guard,xshift=0mm,yshift=1.7cm,align=left] {$qualif=$\\\false} (evalapp);
\draw[sequence,->,rounded corners=5pt] (evalapp) -| (eg1);

\node[ErrorIntermediateEvent,fill=white] (eb) at (9.3,-4.2) {};

\node[task,align=center] (eligcand) at (10.3,-5.3) {\tname{Decide}\\\tname{Eligible}\\\tname{Candidates}};
\draw[sequence,->,rounded corners=5pt] (eb) |- (eligcand);

\node[ExclusiveGateway,draw] (eg_elig) at (12,-5.3) {};
\draw[sequence,->]  (eligcand) --  (eg_elig);

\node[task,align=center] (dwinner) at (12,-2.3) {\tname{Select}\\\tname{Winner}};
\draw[sequence,->,rounded corners=5pt]  (eg_elig) -- node[guard,xshift=1.3cm,yshift=-.1cm,align=left] {$exists\_eligible=$\\\true}  (dwinner);

\node[task,align=center] (awinner) at (12,0) {\tname{Assign}\\\tname{Winner}};
\draw[sequence,->,rounded corners=5pt] (10.93,0) -- (awinner);

\node[ExclusiveGateway,draw] (eg3) at (13.3,0) {};
\draw[sequence,->]  (awinner) --  (eg3);
\draw[sequence,->,rounded corners=5pt]  (dwinner) -|  (eg3);

\node[task,align=center] (offer) at (14.5,0) {\tname{Make}\\\tname{Offer}};
\draw[sequence,->]  (eg3) --  (offer);

\node[EndEvent] (ee2) at (15.9,0) {};
\node[lbl,below of=ee2,align = center] (ee2_name) {};
\draw[sequence,->]  (offer) --  (ee2);

\node[EndEvent] (ee3) at (15.9,-5.3) {};
\node[lbl,below of=ee2,align = center] (ee2_name) {};
\draw[sequence,->]  (eg_elig) --  node[guard,xshift=-.2cm,yshift=-.3cm,align=left] {$exists\_eligible=$\\\false} (ee3);
\end{tikzpicture}
}
\caption{Job hiring process}
\label{fig:ex1}
\end{figure}
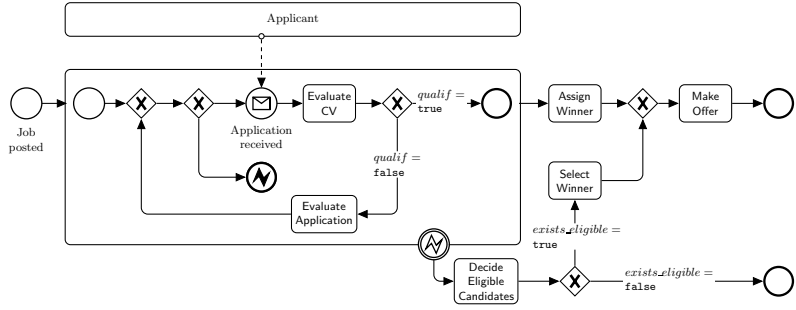
We consider a very slight variant of the example of a job hiring process in a company presented in the paper. 
The human resource (HR) branch of the company stores in its catalog 
database information relevant to the process.  
Specifically, the company's catalog $Cat$ is composed of the following relations:
\begin{compactitem}
\item $JobCategory(\mathit{Jcid}:\sort{jobcatID})$ is used to access different types of jobs that are available in the company ;
\item $User(Uid:\sort{userID},Name:\sort{StringName},Age:\sort{NumAge})$ stores data about users registered to the company website, who might be potentially interested in job positions offered by the company.
\end{compactitem}
To manage information about submitted applications, including data on users, the score they receive after having been interviewed and their eligibility, the company employs repository $Rep$ that consists of one relation $Application(Jcid:\sort{jobcatID},Uid:\sort{userID},Name:\sort{StringName},Age:\sort{NumAge},Score:\sort{NumScore},Eligible:\sort{BoolString})$ that stores all the information. Notice that $\sort{NumScore}$ contains $100$ values in the range
      $(\constant{0},\constant{101})$, where a score from $\constant{1}$ to
  $\constant{100}$ indicates the actual one assigned after evaluating the
  application. For readability, we use the usual predicates $<$, $>$,
  and $=$ to compare variables of type $\sort{NumScore}$: this is syntactic sugar and does not require to introduce rigid predicates in our framework.

Since the job posting is created using a dedicated company's portal, the information related to this posting does not have to be stored persistently and thus can be maintained just for a given case. 
To represent it we use a set of case variables $V_C$, where $jcid:\sort{jobcatID}$ references a job type, $uid:\sort{userID}$ user's identifier together with her name $name:\sort{StringName}$ and age $age:\sort{NumAge}$, a user that wins the position $winner:\sort{userID}$ together with the check of her eligibility $result:\sort{BoolString}$. 
Moreover, we use three more auxiliary variables: $qualif:\sort{Bool}$ identifies whether a currently selected applicant is qualified or not, $exists\_eligible:\sort{Bool}$ indicates successful (or not) termination of the eligible candidate selection process and $tPassed:\sort{StringDate}$ indicates the time passed from the moment when company started receiving applications for the open position.
In the following we described data updates issued by tasks and events in out process model. The execution starts by receiving a new job posting that generates a new posting identifier using the following effect: $\eff(\ename{Job posted}) =\set{\SET\, jcid'=id_j}$. 
As soon the job offer has been published, the company starts a process of receiving and evaluating applications. Such process runs until a qualified candidate is found: nevertheless, if no qualified candidate is found after a non-deterministically assigned deadline has been reached, the process is interrupted anyway. In our case the deadline is modeled using an error event with effect $\eff(ErrorEvent)=\{\SET tPassed'=\constant{1Month}\}$.

Whenever a new application is received, the $\ename{Application received}$ event gets triggered and assigns user data that came together with the application to designated case variables using the following effect:

$$
\eff(\ename{Application received})=
\left\{
 \begin{aligned}
& \SET \,uid'= id_u,\\
& \SET \,name'=n,\\
& \SET \, age'=a\,
 \end{aligned}
 \right\}
 $$

Next, a CV attached to the application undergoes a preliminary evaluation with the sole purpose of detecting a candidate that may be a perfect fit for the position, and thus should immediately win the competition. This is modeled by the $\tname{Evaluate CV}$ task, s.t. $\tpre(\tname{Evaluate CV})=\true$\footnote{Hereinafter we shall avoid putting explicitly trivial (i.e., containing only $\true$) preconditions.} and $\eff(\tname{Evaluate CV})=\{\SET qualif':=o\}$.

If a candidate is not perfectly apt for the position in absentia, we proceed directly to the thorough evaluation of the application followed by the process immediately recording the interview result using the $\tname{Evaluate Application}$ task. This task requires a precondition on data updates, i.e.

$$\tpre({\tname{Evaluate Application}})= 
\left\{
 \begin{aligned}
 & {} (exists\_eligible=\true \land 0<s<101\land y=\true) \lor \\
 &\lor (exists\_eligible=\false \land 80<s<101\land y=\true) \lor \\
 &\lor (exists\_eligible=\false \land 0<s\leq 80\land y=\false)
  \end{aligned}
 \right\}
 $$

 and its update inserts a new tuple to the process repository with a score $s$ such that $0<s<101$: 
$$\eff(\tname{Evaluate Application})=
\left\{
 \begin{aligned}
  & \INSERT (jcid,uid,name,age,s,\nullv)\\
  & \INTO  Application \SETTING exists\_eligible \TO y
 \end{aligned}
 \right\}$$
 
 Notice that the the case variable $exists\_eligible$ becomes $\true$ in case at least an applicant is evaluated with score greater that $80$.

In case a candidate is considered to be perfectly qualified for the position, it immediately gets considered as a winner of the selection process. Such a functionality is carried over by the $\tname{Assign Winner}$ task that simply assigns the winners user identifier to the dedicated case variable: 
$$
\eff(\tname{Assign Winner})=
\left\{
 \begin{aligned}
  & \SET \,winner'=uid, \\
  & \SET \, result'=qualified \\
  & \SET \, tPassed'=LessThan1Month 
 \end{aligned}
 \right\}$$

If the process of the application evaluation has ended up due to the deadline, 
the process runs a task that decides on the eligible candidates among all those that have sent applications. Here as eligible we consider only those candidates whose interview score is greater than 80, whereas others are regarded as not eligible. 
This is done using the \tname{Decide Eligible Candidate} task with the following update:
$$
\eff(\tname{Decide Eligible Candidate})=
\left\{
 \begin{aligned}
& \UPDATE Application(JCID,UID,N,A,S,Elig) \\ 
& \IF S>80 \\ 
& \THEN Application(JCID,UID,N,A,S,'\constant{eligible}') \\ 
& \ELSE Application(JCID,UID,N,A,S,'\constant{noteligible}')
 \end{aligned}
 \right\}$$
Note that the \emph{if-then-else} clause allows us to perform a sort of a bulky update over the repository relation $Application$ by changing the eligibility status of its entries. 

If case there is at least one eligible candidate, she can be selected as a winner. 
This is done by the \tname{Select Winner} task that nondeterministically selects one such candidate from $Application$ (i.e., $\tpre(\tname{Select Winner}) = Application(Jcid,Uid,Name,Age,Score,Eligible)\land Eligible='\constant{eligible}'$) and moves her data to the case variables:

$$
\eff(\tname{Select Winner})=
\left\{
 \begin{aligned}
& \MOVE \,(Jcid, Uid, Name, Age, Eligible) \FROM Application\\ 
& \TO (jcid,uid,name,age, result)\\ 
& \SETTING  tPassed=\constant{1MonthPlus1Week} \\
& \land qualif'=\false \land exists\_eligible'=\nullv
 \end{aligned}
 \right\}$$

Here we also take into account that in order to decide on the eligibility of candidates as well as a winning candidate, the HR staff of the company may require some time. This is duly represented by updating the amount of time passed since the beginning of the candidate selection process. 

At last, when a winning candidate has been selected, the HR office prepares a official offer that is then sent to the winner. In our model this is represented with the \tname{Make Offer} activity that does not issue any updates on the data component (i.e., $\eff(\tname{Make Offer})=\true$).

We analyze four examples of reachability queries, taken from the \mcmt specifications that we tested in our experimental evaluation: we focus on 1-case safety verification. 

The first query expresses that the job hiring process has completed, i.e. it has reached its final state: the tool returns \unsafe (since it exists a sequence of configurations starting from the initial states to the final one), as expected. Formally, we have:
$$
  \begin{array}{@{}l@{}}
   \exists  \typedvar{i}{PI_{index}}
    \left(  \begin{array}{@{}l@{}} lifecycleProcess[i]=\constant{completed} 
     \end{array}
    \right)
  \end{array}
$$

The second query formalizes the situation where, after the evaluation of an application (i.e., EvaluateApplication is completed), there exists at least an applicant with score greater than $0$: the tool returns \unsafe, as expected. Formally, we have:

$$
  \begin{array}{@{}l@{}}
   \exists  \typedvar{i}{PI_{index}}
    \left(  \begin{array}{@{}l@{}} lifecycleEvaluateApplication[i]=\constant{completed} \land \\
     {}\land  Application(Jcid,Uid,Name,Age,Score, Eligible)\land Score>0
     \end{array}
    \right)
  \end{array}
$$

The third query represents the configuration of the system in which a winner has been selected after deadline (i.e., SelectWinner is completed), but the case variable $result$ witnesses that the winner was a not eligible candidate: the tool returns \safe (since this configuration is not reachable from the initial states), as expected. Formally, we have
$$
  \begin{array}{@{}l@{}}
   \exists  \typedvar{i}{PI_{index}}
    \left(  \begin{array}{@{}l@{}} lifecycleSelectWinner[i]=\constant{completed} \land result[i]=\constant{noteligible}
     \end{array}
    \right)
  \end{array}
$$

The final query describes the configuration in which, after the evaluation of an application, there exists an applicant with score greater than $100$: the tool returns \safe, as expected. Formally, we have:
$$
  \begin{array}{@{}l@{}}
   \exists  \typedvar{i}{PI_{index}}
    \left(  \begin{array}{@{}l@{}} lifecycleEvaluateApplication[i]=\constant{completed} \land \\
     {}\land  Application(Jcid,Uid,Name,Age,Score, Eligible)\land Score>100
     \end{array}
    \right)
  \end{array}
$$

All these queries has been checked running MCMT over the running example.

\end{example}

\end{document}